\documentclass[12pt,a4paper]{article}
\usepackage[english]{babel}
\usepackage[numbers,comma,square,compress]{natbib}
\usepackage{amssymb,amsmath,amsfonts,bm,braket,enumerate}

\usepackage{ifpdf}
\ifpdf  % Compiling PDF with pdflatex
 \usepackage[pdftex,a4paper,hmargin=28mm,top=25mm,bottom=26mm]{geometry}
 \usepackage[pdftex]{hyperref}
\else  % Compiling DVI with latex
 \usepackage[dvips,a4paper,hmargin=28mm,top=25mm,bottom=26mm]{geometry}
 \usepackage[dvips]{hyperref}
\fi
\hypersetup{
pdftitle={How unimodular gravity theories differ from general relativity
at quantum level},
pdfauthor={R. Bufalo, M. Oksanen and A. Tureanu},
}

\numberwithin{equation}{section}

\newcommand{\EH}{Einstein-Hilbert }
\newcommand{\R}[1][4]{{}^{(#1)}\!R}
\newcommand{\sR}{\R[3]}

\newcommand{\Kt}{{}^{(2)}\!K}
\newcommand{\pb}[1]{\left\{#1\right\}}
\newcommand{\cG}{\mathcal{G}}
\newcommand{\cC}{\mathcal{C}}
\newcommand{\cH}{\mathcal{H}}
\newcommand{\bcH}{\bar{\cH}}
\newcommand{\tcH}{\tilde{\cH}}
\newcommand{\cK}{\mathcal{K}}

\newcommand{\cM}{\mathcal{M}}
\newcommand{\cD}{\mathcal{D}}
\newcommand{\cU}{\mathcal{U}}
\newcommand{\cN}{\mathcal{N}}
\newcommand{\cB}{\mathcal{B}}
\newcommand{\cF}{\mathcal{F}}
\newcommand{\bn}{\bm{n}}
\newcommand{\pV}{{}_{\perp}V}
\newcommand{\hg}{\hat{g}}
\newcommand{\nn}{\nonumber\\}

\newcommand{\email}[1]{\footnote{E-mail: \href{mailto:#1}{#1}}}

\begin{document}
\begin{center}
\textbf{\Large How unimodular gravity theories differ from\\
\vspace{0.25em} general relativity at quantum level}

\vspace{2em}

{\large
R. Bufalo$^{a,b,}$\email{rbufalo@ift.unesp.br},
M. Oksanen$^{a,}$\email{markku.oksanen@helsinki.fi} and
A. Tureanu$^{a,}$\email{anca.tureanu@helsinki.fi}}\\
\vspace{1em}
\textit{$^{a}$Department of Physics, University of Helsinki,
P.O. Box 64\\ FI-00014 Helsinki, Finland}\\
\vspace{0.5em}
\textit{$^{b}$Instituto de F\'{\i}sica Te\'orica (IFT),
Universidade Estadual Paulista,\\
Rua Dr. Bento Teobaldo Ferraz 271, Bloco II,\\
01140-070 S\~ao Paulo, SP, Brazil}
\end{center}

\vspace{-.5em}

\begin{abstract}
We investigate path integral quantization of two versions of unimodular
gravity. First a fully diffeomorphism-invariant theory is analyzed,
which does not include a unimodular condition on the metric, while still
being equivalent to other unimodular gravity theories at the classical
level. The path integral has the same form as in general relativity
(GR), except that the cosmological constant is an unspecified value of a
variable, and it thus is unrelated to any coupling constant. When the
state of the universe is a superposition of vacuum states, the path
integral is extended to include an integral over the cosmological
constant.
Second, we analyze the standard unimodular theory of gravity, where the
metric determinant is fixed by a constraint. Its path integral differs
from the one of GR in two ways: the metric of spacetime satisfies the
unimodular condition only in average over space, and both the
Hamiltonian constraint and the associated gauge condition have zero
average over space. Finally, the canonical relation between the given
unimodular theories of gravity is established.
\end{abstract}

\section{Introduction}
The idea of unimodular gravity is nearly as old as general relativity
(GR) itself. Originally, Einstein considered the unimodular condition
\cite{Einstein:1916vd},
\begin{equation}\label{unimodcoord}
 \sqrt{-g}=1,
\end{equation}
as a convenient way to partially fix a coordinate system in GR, which
simplifies the calculations in certain situations. Later on, unimodular
gravity has also been considered as an alternative theory of gravity
closely related to GR, which was first suggested in
\cite{Einstein:1919gv}. The definition of unimodular gravity is usually
based on the invariance under a restricted group of diffeomorphisms that
leave the determinant of the metric invariant, so that the determinant
of the metric can be set equal to a fixed scalar density $\epsilon_0$,
\begin{equation}\label{unimodcond}
 \sqrt{-g}=\epsilon_0,
\end{equation}
which provides a fixed volume element in spacetime. We consider a theory
based on the condition \eqref{unimodcond}, and on the associated
restricted group of diffeomorphisms, as the first example of unimodular
gravity.

Fully diffeomorphism-invariant extensions of unimodular gravity
exist as well, which also involve a condition on the determinant of the
metric such that the right-hand side of the condition \eqref{unimodcond}
is replaced with a scalar density field. The most prominent theory
of this kind is the Henneaux-Teitelboim theory \cite{Henneaux:1989zc},
where the unimodular condition sets $\sqrt{-g}$ equal to the divergence
of a vector density field.

It is well known that classically unimodular gravity produces the same
physics as GR with a cosmological constant. The field equation for the
metric is either the traceless Einstein equation or, thanks to the
Bianchi identity, the Einstein equation with a cosmological constant
\cite{Unruh:1988in}. The difference is that the cosmological constant of
unimodular gravity is a constant of integration, rather than a coupling
constant. Since the value of the cosmological constant is unspecified
and unrelated to any coupling constant, problems associated with the
cosmological constant have been reconsidered
(see \cite{Weinberg:1988cp,Padmanabhan:2002ji,Nobbenhuis:2004wn,
Bousso:2007gp,Burgess:2013ara,Padilla:2015aaa} for reviews).

Quantum corrections to the energy-momentum tensor of matter,
$T_{\mu\nu},$ which are of the form $Cg_{\mu\nu}$, where $C$ is a
constant over spacetime, do not contribute to the traceless field
equation for the metric in unimodular gravity. In particular, vacuum
fluctuations in the trace of the energy-momentum tensor of matter do not
affect the metric. This well-known feature of unimodular gravity has
been recently revisited via an explicit calculation of one-loop
corrections \cite{Alvarez:2015pla}. Since a small nonvanishing
cosmological constant is required, the full Einstein equation and an
associated action need to be considered. There the vacuum corrections
are absorbed into the arbitrary cosmological constant, whose value
should be specified experimentally.
However, this does not solve the cosmological constant problem.
Unimodular gravity faces a similar problem with the
renormalization or fine tuning of the cosmological constant as GR
\cite{Weinberg:1988cp,Nobbenhuis:2004wn,Burgess:2013ara,
Padilla:2015aaa}. The expression for the vacuum energy generated by the
quantum fluctuations is highly dependent on the details of the
effective description, in particular on the chosen Wilsonian cut-off
scale \cite{Burgess:2013ara}. Therefore we do not consider the vacuum
energy problem in this paper. Instead we concentrate on the formal
differences between the unimodular gravity theories and GR at the
quantum level. In other words, our treatment assumes that somehow the
observed cosmological constant $\Lambda$ will be stabilized against
vacuum corrections.

Predicting or deriving the observed value of the cosmological constant
is a hard problem as well. A highly speculative but interesting attempt
to address this problem in unimodular gravity has been made in
\cite{Ng:1990rw,Ng:1990xz,Smolin:2009ti}, where an integral over the
cosmological constant was included into the path integral. We will show
how this argument can be derived in a straightforward way, when a new
action for unimodular gravity is introduced. Problems associated with
the given argument are also discussed.

Conventionally, the idea of unimodular gravity has been to impose a
condition on the determinant of the metric, e.g., \eqref{unimodcond}. In
comparison with GR, making the cosmological constant an arbitrary
constant of integration can be regarded as the key feature of
unimodular gravity. In order to achieve it, however, there is no need
to constrain the determinant of the metric. We consider a fully
diffeomorphism-invariant theory (see \eqref{SDUG}), which has
recently appeared in the context of gravity with mimetic dark matter
\cite{Chamseddine:2014vna}, where an additional scalar field
was also included to describe the mimetic matter.
The given theory is no longer unimodular in the sense that there is no
condition on the determinant of the metric, but we will establish how
the theory is canonically related to the conventional unimodular
theories of gravity.

It has been argued that unimodular gravity can offer a new perspective
on the problem of time in quantum gravity and cosmology
\cite{Unruh:1988in,Ng:1990xz,Sorkin:1987cd,Unruh:1989db}. Since the bulk
part of the Hamiltonian of unimodular gravity is nonvanishing, and the
four-volume provides a cosmological time, an analogy of the
Schr\"odinger equation exists, and hence quantum states of the universe
can evolve in terms of a global time. On the other hand, it has been
concluded that unimodular gravity cannot solve the problem of time in
quantum gravity \cite{Kuchar:1991xd}, since the four-volume labels only
equivalence classes of hypersurfaces separated by a zero four-volumes.

Since all versions of unimodular gravity must be classically equivalent
to GR, quantization of each version of unimodular gravity can be
regarded as a potential quantization of GR. Hence it is necessary to
understand how the different versions of unimodular gravity differ from
each other and from GR at the quantum level. The equivalence of GR and
unimodular gravity was recently discussed in \cite{Padilla:2014yea},
concluding that the equivalence can be retained at quantum level when
the UV extension of unimodular gravity is performed appropriately. We
will see that the form of the path integral depends on which version of
unimodular gravity is chosen.

Path integral quantization of the Henneaux-Teitelboim version of
unimodular gravity has been considered previously in
\cite{Smolin:2009ti} (see also \cite{Smolin:2010iq}), where the
unimodular condition was shown to be imposed locally in the quantum
theory. In this paper, we study the path integral quantization for the
two other versions of unimodular gravity discussed above. The results
are compared to both GR and the Henneaux-Teitelboim theory. In the fully
diffeomorphism-invariant theory (see \eqref{SDUG} for action), the path
integral has the same form as the one of GR with a cosmological
constant, but the value of $\Lambda$ is an unspecified constant value of
a variable. Two approaches regarding the interpretation of the
cosmological constant are considered: either (i) the effective value of
the cosmological constant is fixed by the physical boundary conditions
of the path integral, or (ii) the state of the universe is taken as a
superposition of states with different values of $\Lambda$, and
consequently the path integral includes an integral over $\Lambda$. In
the latter approach, we derive the path integral in the form originally
proposed in \cite{Ng:1990rw} (see also \cite{Ng:1990xz,Smolin:2009ti}).
In the theory with a fixed metric determinant (see \eqref{SUG} for
action), the unimodular condition \eqref{unimodcond} is found to be
imposed in average over space, but not locally.

In Sect.~\ref{sec2} we present the different actions of unimodular
gravity which are relevant for this paper, and discuss how the
(classical) actions are related to each other. Section~\ref{sec3} is
devoted to the canonical path integral quantization of the fully
diffeomorphism-invariant theory of unimodular gravity. In
Sect.~\ref{sec4} the same is achieved for the conventional version of
unimodular gravity with a fixed metric determinant. Section~\ref{sec5}
establishes the canonical relation of the theories. The results are
discussed in Sect.~\ref{sec6}.

\section{Three versions of unimodular gravity}\label{sec2}
\subsection{Unimodular gravity with a fixed metric determinant}
\label{sec2.1}
Conventionally, the field equations of unimodular gravity are
obtained from the \EH action under a restricted variation of the metric
$g_{\mu\nu}$ that preserves the determinant of the metric,
\begin{equation}\label{unimodvar}
 \frac{\delta}{\delta g_{\mu\nu}}\sqrt{-g}=0,
\end{equation}
where $g=\det g_{\mu\nu}$. Since the metric transforms under an
infinitesimal diffeomorphism,
\begin{equation}
 x^{\prime \mu}(x)=x^\mu+\xi^\mu(x),
\end{equation}
as
\begin{equation}\label{inDiffg}
 \delta_\xi g_{\mu\nu}=\nabla_\mu\xi_\nu+\nabla_\nu\xi_\mu,
\end{equation}
the unimodular condition \eqref{unimodvar} requires that
\begin{equation}
 \delta\sqrt{-g}=\frac{1}{2}\sqrt{-g}g^{\mu\nu}\delta g_{\mu\nu}=0,
\end{equation}
i.e.,
\begin{equation}\label{divxi}
 \nabla_\mu\xi^\mu=0.
\end{equation}
These transformations are often referred to as transverse
diffeomorphisms or volume-preserving diffeomorphisms. However, the name
transverse diffeomorphisms (TDiff) is sometimes reserved for the
transformations that satisfy the noncovariant condition
$\partial_\mu\xi^\mu=0$ \cite{Buchmuller:1988wx}.
In order to avoid any confusion, we shall refer to the given
transformations \eqref{unimodvar}--\eqref{divxi} as metric
determinant-preserving diffeomorphisms.

One way to define unimodular gravity is to introduce the unimodular
condition \eqref{unimodcond} into \EH action as a constraint multiplied
by a Lagrange multiplier $\lambda$,
\begin{equation}\label{SUG}
\begin{split}
 S_\mathrm{UG}[g_{\mu\nu},\lambda,\Psi]&=\int_{\cM}d^4x\left(
 \frac{\sqrt{-g}R}{\kappa} -\lambda\left( \sqrt{-g} -\epsilon_0 \right)
 \right) \\
 &\quad +\frac{2}{\kappa}\oint_{\partial\cM}d^3x\sqrt{|\gamma|}\cK
 +S_\mathrm{m}[g_{\mu\nu},\Psi],
\end{split}
\end{equation}
where $\epsilon_0$ is a fixed scalar density, such that $\epsilon_0d^4x$
defines a proper volume element, the gravitational coupling constant is
denoted as $\kappa=16\pi G$, and $S_\mathrm{m}$ is the action for
the matter fields (denoted collectively by $\Psi$) which are coupled to
the metric in the same way as in GR. In the surface integral over the
boundary $\partial\cM$ of spacetime, $\gamma$ is the determinant of the
induced metric on $\partial\cM$, and $\cK$ is the trace of the extrinsic
curvature of the boundary. The boundary term is included as in GR, so
that the variational principle for the action is well defined without
imposing boundary conditions on the derivatives of the
metric.\footnote{When we write about boundary conditions without
specifying their nature in the canonical formalism, we refer to both the
initial conditions and the conditions on the spatial boundary. Likewise
in Lagrangian formalism we refer to conditions on the boundary of
spacetime.}
The full diffeomorphism invariance of GR is lost due to the presence of
the fixed volume element $\epsilon_0d^4x$.
The action \eqref{SUG} is invariant under the metric
determinant-preserving diffeomorphisms. We shall refer to the theory
defined by \eqref{SUG} simply as unimodular gravity (UG).

% Let us obtain the field equations.
An unrestricted variation of $g^{\mu\nu}$ gives the Einstein equation
\begin{equation}
 R_{\mu\nu}-\frac{1}{2}Rg_{\mu\nu}+\frac{\kappa}{2}\lambda g_{\mu\nu}
 =\frac{\kappa}{2} T_{\mu\nu}.\label{Efe0}
\end{equation}
The variation of $\lambda$ gives the unimodular condition
\eqref{unimodcond}. The field equations for matter fields are identical
to those of GR. The unimodular condition \eqref{unimodcond} ensures that
\eqref{unimodvar} holds. The energy-momentum tensor of matter is defined
in \eqref{Efe0} as usual, $T_{\mu\nu}=-\frac{2}{\sqrt{-g}}
\frac{\delta S_\mathrm{m}}{\delta g^{\mu\nu}}$.
We assume that the action for matter is diffeomorphism-invariant, so
that energy-momentum is conserved, $\nabla^\nu T_{\mu\nu}=0$.
Then we take the divergence of \eqref{Efe0} and obtain
\begin{equation}
 \nabla_\mu\lambda=\nabla^\nu T_{\mu\nu}
 -\frac{2}{\kappa}\left( \nabla^\nu R_{\mu\nu}-\frac{1}{2}\nabla_\mu R
\right)=0,
\end{equation}
where the (contracted) Bianchi identity is used. Thus we see that
$\lambda$ is fixed as a constant of integration, which we denote as
$\lambda=\frac{2}{\kappa}\Lambda$, where $\Lambda$ is the cosmological
constant. Inserting this into the field equation
\eqref{Efe0} gives
\begin{equation}\label{Efe}
 R_{\mu\nu}-\frac{1}{2}Rg_{\mu\nu}+\Lambda g_{\mu\nu}
 =\frac{\kappa}{2} T_{\mu\nu}.
\end{equation}
Compared to GR the only difference is that we are restricted to use
coordinate systems that satisfy \eqref{unimodcond}. In GR, the condition
\eqref{unimodcond} can always be satisfied locally by choosing the
inertial coordinates. Then every coordinate system obtained via
metric determinant-preserving diffeomorphisms satisfies
\eqref{unimodcond} as well.

\subsection{Fully diffeomorphism-invariant unimodular gravity}
\label{sec2.2}
Extensions of unimodular gravity with full diffeomorphism invariance
have been proposed as well. The most prominent theory is defined by the
Henneaux-Teitelboim (HT) action \cite{Henneaux:1989zc} (see
\cite{Fiol:2008vk} regarding the boundary surface term),
\begin{equation}\label{SHT}
\begin{split}
 S_\mathrm{HT}[g_{\mu\nu},\lambda,\tau^\mu,\Psi]&=\int_{\cM} d^4x
 \left( \frac{\sqrt{-g}R}{\kappa}
 -\lambda\left( \sqrt{-g}-\partial_\mu\tau^\mu \right) \right)\\
 &\quad+\oint_{\partial\cM}d^3x\left( \frac{2}{\kappa}\sqrt{|\gamma|}\cK
 -\lambda r_\mu\tau^\mu \right)
 +S_\mathrm{m}[g_{\mu\nu},\Psi],
\end{split}
\end{equation}
where $\tau^\mu$ is a vector density and $r_\mu$ is the
outward-pointing unit normal to the boundary $\partial\cM$.
The field equations consists of the Einstein equation \eqref{Efe0}, the
equation for the cosmological constant variable,
\begin{equation}\label{nablalambda}
 \nabla_\mu\lambda=0,
\end{equation}
a (fully diffeomorphism-invariant) unimodular condition,
\begin{equation}\label{unimodcond.tau}
 \sqrt{-g}=\partial_\mu\tau^\mu,
\end{equation}
and standard field equations for matter.

The HT action \eqref{SHT} can indeed be derived from the UG action
\eqref{SUG} via parameterization of the spacetime coordinates
\cite{Kuchar:1991xd}. Parametrization of coordinates in a mechanical
system is a well known method for obtaining a reparameterization
invariant action (see \cite{Lanczos:1970var} for a review).
Parametrization of field theories was introduced later (see
\cite{Dirac:LQM} for a description). We treat the coordinates of the
action \eqref{SUG} as four independent scalar variables $X^\alpha(x)$
that depend on the actual coordinates $x^\mu$. One can think of this as
a transformation $x^\alpha\rightarrow X^\alpha(x)$. The \EH and matter
parts of the action \eqref{SUG} are invariant under such transformation,
but the part with a fixed volume element is not invariant, since it
transforms as  $\int d^4x\epsilon_0\lambda\rightarrow\int
d^4x\epsilon_0\lambda|\partial_\mu X^\alpha|$, where $|\partial_\mu
X^\alpha|$ is the Jacobian determinant of the transformation. When we
identify a vector density as $\tau^\mu=4!\epsilon_0
\delta^{[\mu}_\alpha\delta^\nu_\beta\delta^\rho_\gamma
\delta^{\sigma]}_\delta X^\alpha\partial_\nu X^\beta \partial_\rho
X^\gamma\partial_\sigma X^\delta$, we obtain the HT action \eqref{SHT}.
It is clear that the HT theory is classically equivalent to the UG
theory \eqref{SUG}. However, differences are expected to arise upon
quantization.

We consider an alternative action that is fully diffeomorphism-invariant
and retains the classical equivalence with the other unimodular
theories, in particular with \eqref{SUG} and \eqref{SHT}.
The action has been studied in the context of gravity with mimetic dark
matter \cite{Chamseddine:2014vna}, where an additional scalar field was
also included to describe the mimetic matter.
The action is written (without the scalar field) as
\begin{equation}\label{SDUG}
\begin{split}
 S_\mathrm{DUG}[g_{\mu\nu},\lambda,V^\mu,\Psi]&=\int_{\cM} d^4x\sqrt{-g}
 \left( \frac{R}{\kappa} -\lambda -V^\mu\nabla_\mu\lambda \right)\\
 &\quad+\frac{2}{\kappa}\oint_{\partial\cM}d^3x\sqrt{|\gamma|}\cK
 +S_\mathrm{m}[g_{\mu\nu},\Psi],
\end{split}
\end{equation}
where the variable $V^\mu$ is a vector field. We shall refer to this
theory as the fully diffeomorphism-invariant unimodular gravity (DUG).
The action \eqref{SDUG} is arguably the most transparent definition of
such a theory. The action \eqref{SDUG} consists of the \EH action with a
variable cosmological constant $\lambda$, and a constraint term for
$\lambda$. The vector field $V^\mu$ acts as a Lagrange multiplier that
ensures $\nabla_\mu\lambda$ is zero in every direction, and thus
$\lambda$ is a constant. The field equations consists of the Einstein
equation \eqref{Efe0} for the metric, the equation \eqref{nablalambda}
for the cosmological constant variable $\lambda$, an equation for the
auxiliary vector field
\begin{equation}\label{divV}
 \nabla_\mu V^\mu=1,
\end{equation}
and standard field equations for matter.
The unimodular condition on the metric determinant, \eqref{unimodcond}
or \eqref{unimodcond.tau},  has been replaced with the condition
\eqref{divV} on the vector field. The vector field does not contribute
to the Einstein equation due to Eq.~\eqref{nablalambda}. In
Sect.~\ref{sec3}, we will show how the vector field can be eliminated
from the Hamiltonian formulation while the canonical representation of
diffeomorphism invariance is retained.

It is obvious that the DUG action \eqref{SDUG} is closely related to the
HT action \eqref{SHT}. An integration by parts in the term $\int d^4x
\lambda\partial_\mu\tau^\mu$ of the HT action, followed by a replacement
of the vector density variable with a vector field variable,
$\tau^\mu=\sqrt{-g}V^\mu$, gives the action \eqref{SDUG}. Hence it is
clear that these theories are equivalent classically. However, the path
integral for the action \eqref{SDUG} will be shown to differ from the HT
case significantly due to the different choice of variable.

The field equations for both the HT and the DUG theories are invariant
under the shift
\begin{equation}
 T_{\mu\nu}\rightarrow T_{\mu\nu}
 +Cg_{\mu\nu},\quad \lambda\rightarrow\lambda+C,
\end{equation}
where $C$ is a constant. Hence quantum corrections to the trace of the
energy-momentum tensor are absorbed into the variable $\lambda$, whose
value is an arbitrary constant. The variable $\lambda$ will be shown to
remain constant at quantum level in Sect.~\ref{sec3}.

There exist more versions of unimodular gravity in addition to the three
theories discussed above; see, for example,
\cite{Smolin:2009ti,Padilla:2014yea} for other actions. In this paper we
will concentrate on the three theories defined by \eqref{SUG},
\eqref{SHT} and \eqref{SDUG}.

\section{Quantization of the fully diffeomorphism-invariant unimodular
gravity}\label{sec3}
\subsection{Arnowitt-Deser-Misner decomposition of the
action}\label{sec3.1}
Spacetime is assumed to admit a foliation to a union of nonintersecting
spacelike hypersurfaces. The hypersurfaces $\Sigma_t$ are labelled by a
scalar $t$ that is constant across each hypersurface. The
future-pointing unit normal to $\Sigma_t$ is denoted by $n^\mu$. The
so-called direction of time vector $t^\mu$ satisfies $t^\mu\nabla_\mu
t=1$. The metric $g_{\mu\nu}$ has the signature $(-,+,+,+)$, and hence
$n_\mu n^\mu=-1$. Each hypersurface is described by the induced metric
on $\Sigma_t$,
\begin{equation}
 h_{\mu\nu}=g_{\mu\nu}+n_\mu n_\nu,
\end{equation}
and by the extrinsic curvature tensor
\begin{equation}\label{Kmunu}
 K_{\mu\nu}=\nabla_\mu n_\nu+n_\mu a_\nu,
\end{equation}
where we have defined the acceleration vector of Eulerian observers as
\begin{equation}\label{amu}
 a_\mu=n^\nu\nabla_\nu n_\mu.
\end{equation}

Then we introduce the Arnowitt-Deser-Misner (ADM) variables.
The scalar $t$ is taken as the time coordinate. The unit normal to
$\Sigma_t$ is written as
\begin{equation}
 n^0=\frac{1}{N},\quad n^i=-\frac{N^i}{N},
\end{equation}
where $N$ is the lapse variable and $N^i$ is the shift vector on
$\Sigma_t$. Latin indices ($i,j,\ldots$) range from 1 to 3. Now
the metric takes the form
\begin{equation}
 g_{00}=-N^2+N_iN^i,\quad g_{0i}=N_i,\quad g_{ij}=h_{ij},
\end{equation}
where $N_i=h_{ij}N^j$. The extrinsic curvature is written as
\begin{equation}\label{Kij}
 K_{ij}=\frac{1}{2N}\left( \partial_th_{ij}-D_iN_j-D_jN_i \right),
\end{equation}
where $D$ is the covariant derivative that is compatible with the metric
$h_{ij}$ on $\Sigma_t$, and $h^{ij}$ is the inverse metric,
$h_{ij}h^{jk}=\delta_i^{\ k}$. The trace of extrinsic curvature is
denoted by $K=h^{ij}K_{ij}$.

The action is decomposed as follows. The metric determinant is given by
\begin{equation}
 \sqrt{-g}=N\sqrt{h},
\end{equation}
where $h=\det h_{ij}$. The scalar curvature is written as
\begin{equation}\label{R.ADM}
 R=K_{ij}\cG^{ijkl}K_{kl}+\sR+2\nabla_\mu(n^\mu K-a^\mu),
\end{equation}
where the De Witt metric is defined as
\begin{equation}
 \cG^{ijkl}=\frac{1}{2}(h^{ik}h^{jl}+h^{il}h^{jk})-h^{ij}h^{kl}
\end{equation}
and $\sR$ is the (intrinsic) scalar curvature of $\Sigma_t$.
The last term in \eqref{R.ADM} is a total derivative which contributes a
boundary term into the action.

The vector field is decomposed into components tangent and
normal to $\Sigma_t$ as
\begin{equation}\label{Vdec}
 V^\mu=\pV^\mu-n^\mu V_{\bn},
\end{equation}
where
\begin{equation}
  \pV^\mu=h^\mu_{\ \nu}V^\nu,\quad V_{\bn}=n_\mu V^\mu,
\end{equation}
and the projection operator onto $\Sigma_t$ is defined as
\begin{equation}\label{projector}
 h^\mu_{\ \nu}=\delta^\mu_{\ \nu}+n^\mu n_\nu.
\end{equation}
The gravitational part of the action \eqref{SDUG} is written in ADM form
as
\begin{multline}\label{SDUG.ADM}
 S_\mathrm{DUG}[N,N^i,h_{ij},\lambda,V_{\bn},V^i,\Psi]=\int dt
 \int_{\Sigma_t}d^3x N\sqrt{h}\bigg[
 \frac{1}{\kappa}\left( K_{ij}\cG^{ijkl}K_{kl}+\sR \right)\\
 -\lambda +V_{\bn}\nabla_n\lambda -V^i\partial_i\lambda \bigg]
 +S_{\cB} +S_\mathrm{m}[g_{\mu\nu},\Psi],
\end{multline}
where we denote $V^i=\pV^i$ and
\begin{equation}
 \nabla_n\lambda=\frac{1}{N}\left( \partial_t\lambda
 -N^i\partial_i\lambda \right),
\end{equation}
and the boundary contribution $S_{\cB}$ is given as in GR,
\begin{equation}\label{ScB}
 S_{\cB}=\frac{2}{\kappa}\int_{\cB}d^3x\sqrt{-\gamma}\left( \cK
 +r_\mu n^\mu K -r_\mu a^\mu \right),
\end{equation}
where $\cB$ is the timelike part of the boundary $\partial\cM$. The
surface $\cB$ is foliated to a union of two-dimensional surfaces
$\cB_t$, which come from the intersection of $\Sigma_t$ and $\cB$.
When the hypersurfaces $\cB$ and $\Sigma_t$ are orthogonal, the surface
term \eqref{ScB} can be written as \cite{Hawking:1995fd}
\begin{equation}
 S_{\cB}=\frac{2}{\kappa}\int dt\int_{\cB_t}d^2xN\sqrt{\sigma}\Kt,
\end{equation}
where $\sigma$ is the determinant of the induced metric on $\cB_t$, and
$\Kt$ is the trace of the extrinsic curvature of $\cB_t$ in $\Sigma_t$.

\subsection{Hamiltonian analysis}\label{sec3.2}
Hamiltonian analysis of unimodular gravity in its different forms has
been considered in several papers \cite{Henneaux:1989zc,Unruh:1988in,
Smolin:2009ti,Unruh:1989db,Kuchar:1991xd,Smolin:2010iq,Bombelli:1991jj,
Kluson:2014esa}. Since the action \eqref{SDUG} differs from the
previous theories by lacking a unimodular condition and involving the
vector field, we present a detailed Hamiltonian analysis.

\subsubsection{Hamiltonian and constraints}
We shall obtain the Hamiltonian and the full set of constraints for the
action \eqref{SDUG}. Here we consider pure gravity, since the matter
sector is identical to that of GR, and in the end we include matter
into the path integral in Sect.~\ref{sec3.3}.

First we introduce the canonical momenta $\pi_N$, $\pi_i$,
$\pi^{ij}$, $p_\lambda$, $p_i$ and $p_{\bn}$ conjugate to $N$, $N^i$,
$h_{ij}$, $\lambda$, $V^i$ and $V_{\bn}$, respectively.
Since the action \eqref{SDUG.ADM} is independent of the time derivatives
of the variables $N$, $N^i$, $V_{\bn}$ and $V^i$, their canonically
conjugated momenta are primary constraints:
\begin{equation}
 \pi_N\approx0,\quad \pi_i\approx0,\quad p_{\bn}\approx0,\quad
p_i\approx0.
\end{equation}
In addition, the definition of the momentum conjugate to $\lambda$
% \begin{equation}
%  p_\lambda=\frac{\delta S_\mathrm{DUG}}{\delta(\partial_t\lambda)}
%  =\sqrt{h}V_{\bn},
% \end{equation}
implies the primary constraint
\begin{equation}\label{Cbn}
 \cC_\lambda=p_\lambda-\sqrt{h}V_{\bn}\approx0.
\end{equation}
The momentum conjugate to the metric $h_{ij}$ is defined as
\begin{equation}
 \pi^{ij}%=\frac{\delta S_\mathrm{DUG}}{\delta(\partial_th_{ij})}
 =\frac{\sqrt{h}}{\kappa}\cG^{ijkl}K_{kl}.
\end{equation}

The Hamiltonian is obtained as
\begin{equation}\label{HDUG}
 H=\int_{\Sigma_t}d^3x\left( N\cH_T+N^i\cH_i +v_N\pi_N+v_N^i\pi_i
 +v_\lambda \cC_\lambda +v_{\bn}p_{\bn} +v^ip_i \right)
 +H_{\cB_t},
\end{equation}
where the so-called super-Hamiltonian and supermomentum are
defined as
\begin{equation}\label{cHT}
  \cH_T=\frac{\kappa}{\sqrt{h}}\pi^{ij}\cG_{ijkl}\pi^{kl}
 -\frac{\sqrt{h}}{\kappa}\sR +\sqrt{h}\lambda
 +\sqrt{h}V^i\partial_i\lambda
\end{equation}
and
\begin{equation}\label{cHi}
 \cH_i=-2h_{ij}D_k\pi^{jk}+\partial_i\lambda p_\lambda,
\end{equation}
respectively, where we introduced the inverse De Witt metric as
\begin{equation}
 \cG_{ijkl}=\frac{1}{2}(h_{ik}h_{jl}+h_{il}h_{jk})
 -\frac{1}{2}h_{ij}h_{kl},
\end{equation}
and $v_N,v_N^i,v_\lambda,v^i,v_{\bn}$ are unspecified Lagrange
multipliers for the primary constraints.
Regarding the surface terms, the analysis follows the standard set
by \cite{Hawking:1995fd}.
The surface term in the Hamiltonian \eqref{HDUG} is obtained as
\begin{equation}\label{HBt}
 H_{\cB_t}=-\frac{2}{\kappa}\int_{\cB_t}d^2xN\sqrt{\sigma}\Kt
 +2\int_{\cB_t}d^2xN^i h_{ij}r_k\pi^{jk},
\end{equation}
which is the same expression as in GR. The physical Hamiltonian is
defined with respect to a chosen reference background as
$H_\mathrm{phys}=H-H_\mathrm{ref}$, where the Hamiltonian of the
background is denoted as $H_\mathrm{ref}$. The total gravitational
energy of the system is the value of the physical Hamiltonian. The
surface term \eqref{HBt} is given in a generic form that produces the
correct expression of total gravitational energy for different
reference backgrounds \cite{Hawking:1995fd}.

We must ensure that every constraint is preserved under time evolution
that is generated by the Hamiltonian \eqref{HDUG}. The preservation of
$\pi_N\approx0$ is ensured by the Hamiltonian constraint
\begin{equation}
 \cH_T\approx0
\end{equation}
and the preservation of $\pi_i\approx0$ is ensured by the momentum
constraint
\begin{equation}
 \cH_i\approx0.
\end{equation}
We can extend the momentum constraint \eqref{cHi} with a term that is
proportional to the primary constraint $p_{\bn}$ so that the momentum
constraint generates spatial diffeomorphisms on $\Sigma_t$ for
all the variables that are involved in the constraints.\footnote{We do
not need to include a generator for the variables $V^i$ and $p_i$, since
the $V^i$-dependent term in the Hamiltonian constraint \eqref{cHT} is
proportional to the constraint \eqref{cCi} found below.} For that reason
we redefine
\begin{equation}
 \cH_i=-2h_{ij}D_k\pi^{jk}+\partial_i\lambda p_\lambda
 +\partial_iV_{\bn}p_{\bn}\approx0.
\end{equation}
It is useful to define global (smeared) versions of these constraints
for calculational purposes:
\begin{equation}\label{HMC}
 \cH_T[\xi]=\int_{\Sigma_t}d^3x\xi\cH_T,\quad
 \Phi[\chi^i]=\int_{\Sigma_t}d^3x \chi^i\cH_i,
\end{equation}
where $\xi$ and $\chi^i$ are functions on $\Sigma_t$.
The preservation of the constraint $p_i\approx0$,
\begin{equation}
 \partial_tp_i=\pb{p_i,H}\approx-N\sqrt{h}\partial_i\lambda\approx0,
\end{equation}
is ensured by introducing a new constraint,
\begin{equation}\label{cCi}
 \cC_i=\partial_i\lambda\approx0.
\end{equation}
This constraint implies that $\lambda$ is a constant across $\Sigma_t$.
We define the smeared form of $\cC_i$ as
\begin{equation}\label{cC}
 \cC[\chi^i]=\int_{\Sigma_t}d^3x\chi^i\partial_i\lambda.
\end{equation}
This constraint is included into the Hamiltonian with a Lagrange
multiplier as $\cC[v_\lambda^i]$.
The preservation of the constraint $\cC_\lambda\approx0$,
\begin{equation}
 \partial_t\cC_\lambda=\pb{\cC_\lambda,H}\approx
 \pb{\cC_\lambda,\cH_T[N]}-\sqrt{h}v_{\bn}
 +\pb{\cC_\lambda,\cC[v_\lambda^i]} \approx0,
\end{equation}
is ensured by fixing the Lagrange multiplier $v_{\bn}$ of the
constraint $p_{\bn}$ as
\begin{equation}
 v_{\bn}=-N+\frac{\kappa}{2} N\frac{h_{ij}\pi^{ij}}{\sqrt{h}}V_{\bn}
 +\frac{1}{\sqrt{h}}\partial_iv_\lambda^i.
\end{equation}
The preservation of the constraint $p_{\bn}\approx0$,
\begin{equation}
 \partial_tp_{\bn}=\pb{p_{\bn},H}\approx\sqrt{h}v_\lambda\approx0,
\end{equation}
is ensured by fixing the Lagrange multiplier $v_\lambda$ of the
constraint $\cC_\lambda$ as
\begin{equation}\label{vlambda}
 v_\lambda=0.
\end{equation}
Since the constraint $\cC_i$ is included into the Hamiltonian with a
Lagrange multiplier, we can simplify the system by redefining $\cH_T$
without the part that is proportional to $\cC_i$. Now the total
Hamiltonian is written as
\begin{equation}\label{HDUG2}
 H=\int_{\Sigma_t}d^3x\left( N\cH_T'+N^i\cH_i +v_N\pi_N+v_N^i\pi_i
 +v^ip_i+v_\lambda^i\cC_i' \right) +H_{\cB_t},
\end{equation}
where we have defined the constraints
\begin{align}
 \cH_T'&=\cH_T-p_{\bn}+\frac{\kappa}{2}\frac{h_{ij}\pi^{ij}}{\sqrt{h}}
 V_{\bn}p_{\bn}\approx0,\label{cHT'}\\
 \cH_T&=\frac{\kappa}{\sqrt{h}}\pi^{ij}\cG_{ijkl}\pi^{kl}
 -\frac{\sqrt{h}}{\kappa}\sR +\sqrt{h}\lambda\approx0,\nn
 \cC_i'&=\cC_i-\partial_i\left( \frac{p_{\bn}}{\sqrt{h}} \right)
%  =\partial_i\left( \lambda-\frac{p_{\bn}}{\sqrt{h}} \right)
 \approx0, \label{cCi'}
\end{align}
and $v_N,v_N^i,v^i,v_\lambda^i$ are unspecified Lagrange multipliers.

What remains to be established is the preservation of the secondary
constraints $\cH_T$, $\cH_i$ and $\cC_i$ under time evolution. The
constraints $\cH_T$, $\cH_i$, $\cC_i$ have vanishing Poisson bracket
with $p_{\bn}$. The constraints $\cH_T$, $\cH_i$, $\cC_i$ satisfy the
following algebra:
\begin{align}
 \pb{\cH_T[\xi],\cH_T[\eta]}&=\int_{\Sigma_t}d^3x
 (\xi\partial_i\eta -\eta\partial_i\xi)h^{ij}(\cH_j -p_\lambda\cC_j
 -\partial_jV_{\bn}p_{\bn}),\nn
 \pb{\Phi[\chi^i],\cH_T[\xi]}&=\cH_T[\chi^i\partial_i\xi],\nn
 \pb{\Phi[\chi^i],\Phi[\psi^j]}&=\Phi[\chi^j\partial_j\psi^i
 -\psi^j\partial_j\chi^i],\nn
 \pb{\cH_T[\xi],\cC[\chi^i]}&=0,\nn
 \pb{\Phi[\chi^i],\cC[\eta^j]}&=\cC[\chi^i\partial_j\eta^j].
 \label{scalgebra}
\end{align}
The first three Poisson brackets in \eqref{scalgebra} are the familiar
relations found in GR. The constraint in the right-hand side of the
first Poisson bracket is just the momentum constraint of GR,
$\cH_i-p_\lambda\cC_i-\partial_iV_{\bn}p_{\bn}=-2h_{ij}D_k\pi^{jk}
\approx0$. The last two Poisson brackets tell that $\cC_i$ is preserved
in time and that it transforms as a vector density under the spatial
diffeomorphisms generated by the momentum constraint. The constraints
$\pi_N$, $\pi_i$, $p_i$ have vanishing Poisson bracket with every
constraint. Thus all the constraints are preserved under time evolution.

We can now see that all the constraints ($\cH_T',\cH_i,\pi_N,\pi_i,
p_i,\cC_i'$) in the Hamiltonian \eqref{HDUG2} are first class
constraints. The Lagrange multipliers in the Hamiltonian \eqref{HDUG2}
remain unspecified, until they are determined as a part of the gauge
fixing procedure. In addition, $p_{\bn}\approx0$ and
$\cC_\lambda\approx0$ are the second class constraints.

In order to clarify the nature of the constraint $\cC_i$, it is useful
to decompose the variables $\lambda,p_\lambda$ as follows:
\begin{equation}\label{lambdadec}
\begin{split}
 \lambda(t,x)&=\lambda_0(t)+\bar{\lambda}(t,x),\\
 p_\lambda(t,x)&=\frac{\sqrt{h}}{\int_{\Sigma_t}d^3x\sqrt{h}}
 p_\lambda^0(t)+\bar{p}_\lambda(t,x),
\end{split}
\end{equation}
where the zero modes describe the time dependent averages of $\lambda$
and $p_\lambda$ over space,
\begin{equation}\label{lambdazero}
 \lambda_0(t)=\frac{1}{\int_{\Sigma_t}d^3x\sqrt{h}}
 \int_{\Sigma_t}d^3x\sqrt{h}\lambda(t,x),\quad
 p_\lambda^0(t)=\int_{\Sigma_t}d^3x p_\lambda(t,x),
\end{equation}
and the barred components have vanishing average values over space,
\begin{equation}\label{noaverage}
 \int_{\Sigma_t}d^3x\sqrt{h} \bar{\lambda}(t,x)=0,\quad
 \int_{\Sigma_t}d^3x\bar{p}_\lambda(t,x)=0.
\end{equation}
If the space $\Sigma_t$ is infinite, the definition of the zero
modes \eqref{lambdazero} has to be specified more precisely. For
example, in the asymptotically flat case, the spatial integrals would be
defined up to a finite radius $r$ in the asymptotic region, and finally
the limit $r\rightarrow\infty$ would be taken. In the definition of
$\lambda_0$ the two infinite integrals cancel out, since the asymptotic
value of $\lambda$ must be a constant, so that the average value
$\lambda_0$ remains finite. The momentum $p_\lambda$ can be defined to
have such an asymptotic behavior that the definition of its zero mode
remains finite.
Other scalar fields or scalar densities can be decomposed in a similar
way. The zero modes satisfy the canonical Poisson bracket
\begin{equation}
 \pb{\lambda_0,p_\lambda^0}=1,
\end{equation}
while the average free components satisfy
\begin{equation}
 \pb{\bar{\lambda}(x),\bar{p}_\lambda(y)}=\delta(x-y)
 -\frac{\sqrt{h}(y)}{\int_{\Sigma_t}d^3z\sqrt{h}},
\end{equation}
and the Poisson brackets between zero modes and average free components
are zero
\begin{equation}
 \pb{\lambda_0,\bar{p}_\lambda(x)}=0,\quad
 \pb{\bar{\lambda}(x),p_\lambda^0}=0.
\end{equation}
When $\lambda$ is decomposed, the constraint \eqref{cCi} can be replaced
with a local constraint
\begin{equation}\label{barlambda}
 \bar{\lambda}\approx0,
\end{equation}
since $\partial_i\lambda=\partial_i\bar{\lambda}=0$ implies that
$\bar{\lambda}$ is constant over space and the zero average condition
\eqref{noaverage} requires that constant to be zero. The corresponding
first class constraint \eqref{cCi'} is replaced with
\begin{equation}\label{barcC'}
 \bar{\cC}'=\bar{\lambda}-\overline{\left( \frac{p_{\bn}}{\sqrt{h}}
 \right)} \approx0.
%  =\lambda-\lambda_0 -\frac{p_{\bn}}{\sqrt{h}}
%  +\frac{\int_{\Sigma_t}d^3xp_{\bn}}{\int_{\Sigma_t}d^3x\sqrt{h}}.
\end{equation}
where the overline denotes a component whose integral over space
vanishes. The purpose of the decomposition \eqref{lambdadec} of the
cosmological variable is to separate the perturbative component
$\bar\lambda$ that vanishes due to the constraint \eqref{cCi}. The
average component $\lambda_0$ is left unconstrained.

The total Hamiltonian \eqref{HDUG2} is rewritten as
\begin{equation}\label{HDUG3}
 H=\int_{\Sigma_t}d^3x\left( N\cH'_T+N^i\cH_i +v_N\pi_N+v_N^i\pi_i
 +\bar{v}_\lambda\bar{\cC}'+v^ip_i \right) +H_{\cB_t},
\end{equation}
where the variable $\lambda$ in the Hamiltonian constraint \eqref{cHT'}
is replaced with its zero mode $\lambda_0$.
Next we consider gauge fixing and simplification of the Hamiltonian via
elimination of some variables.

\subsubsection{Gauge fixing and the second class constraints}
Each of the first class constraints generates a gauge transformation.
The constraint $p_i$ generates a gauge transformation of the vector
$V^i$ as
\begin{equation}
 \delta V^i=\pb{V^i,\int_{\Sigma_t}d^3x\varepsilon^ip_i}=\varepsilon^i,
\end{equation}
where $\varepsilon^i$ is an infinitesimal gauge parameter. This means
that $V^i$ can be fixed throughout spacetime as a gauge choice. We
choose the gauge fixing condition as $V^i=0$.
We can further simplify the system by considering the gauge symmetry
that is associated with the constraint \eqref{barcC'}. The constraint
\eqref{barcC'} generates the transformation of the average free
momentum $\bar{p}_\lambda$,
\begin{equation}
 \delta\bar{p}_\lambda=\pb{\bar{p}_\lambda,
 \int_{\Sigma_t}d^3x\bar{\varepsilon}\bar{\cC}'}
 =-\bar{\varepsilon},
\end{equation}
where the infinitesimal gauge parameter $\bar{\varepsilon}$ is a now a
scalar density whose integral over $\Sigma_t$ vanishes. Equivalently,
the constraint \eqref{cCi'} generates the transformation
\begin{equation}
 \delta \bar{p}_\lambda=\pb{\bar{p}_\lambda,
 \int_{\Sigma_t}d^3x\varepsilon^i\cC_i'}
 =\partial_i\varepsilon^i,
\end{equation}
where the integral of the component of the infinitesimal gauge parameter
$\varepsilon^i$ in the direction of the outward-pointing unit normal
$r_i$ to the boundary of $\Sigma_t$ is zero, so that
$\int_{\Sigma_t}d^3x \partial_i\varepsilon^i
=\int_{\cB_t}d^2xr_i\varepsilon^i=0$.
The corresponding gauge freedom can be fixed by setting
$\bar{p}_\lambda=0$.

Now we have the set of second class constraints $(\cC_\lambda,p_{\bn},
V^i,p_i,\bar{\lambda},\bar{p}_\lambda)$. The second class
constraints can be set to zero strongly, if we replace the Poisson
bracket with the Dirac bracket. In this case the Dirac bracket is equal
to the Poisson bracket. Then we can eliminate six canonical variables
$(V_{\bn},p_{\bn},V^i,p_i,\bar{\lambda},\bar{p}_\lambda)$ by using the
constraints
\begin{align}
 V_{\bn}&=\frac{p_\lambda}{\sqrt{h}},\quad p_{\bn}=0,\nn
 V^i&=0,\quad p_i=0,\nn
 \bar{\lambda}&=0,\quad \bar{p}_\lambda=0.
\end{align}
The Hamiltonian \eqref{HDUG3} is written as
\begin{equation}\label{HV4}
 H=\int_{\Sigma_t}d^3x\left( N\cH_T+N^i\cH_i +v_N\pi_N+v_N^i\pi_i
 \right) +H_{\cB_t},
\end{equation}
where
\begin{equation}\label{cHT2}
 \cH_T=\frac{\kappa}{\sqrt{h}}\pi^{ij}\cG_{ijkl}\pi^{kl}
 -\frac{\sqrt{h}}{\kappa}\sR +\sqrt{h}\lambda_0\approx0,
\end{equation}
and
\begin{equation}\label{cHi2}
 \cH_i=-2h_{ij}D_k\pi^{jk}\approx0.
\end{equation}
This is the Hamiltonian of GR with a time dependent cosmological
constant $\lambda_0$. However, it is evident that $\lambda_0$ is a
constant in time as well, since the Hamiltonian is independent of
$p_\lambda^0$,
\begin{equation}
 \partial_t\lambda_0=\pb{\lambda_0,H}=0.
\end{equation}
The value of $\lambda_0$ is set as a part of the initial value data on
the initial Cauchy surface, say $\Sigma_0$ at $t=0$. The momentum
$p_\lambda^0$ evolves monotonically,
$\partial_t p_\lambda^0=-N\sqrt{h}$, and it is not involved in the
actual dynamics of the system. The physical degrees of freedom consist
of the two standard modes of gravity, plus the nondynamical zero mode
that provides the cosmological constant.

Gauge fixing the diffeomorphism invariance is done exactly as in GR by
introducing appropriate gauge conditions for the generators
$\cH_T,\cH_i,\pi_N,\pi^i$.

\subsection{Path integral}\label{sec3.3}
\subsubsection{Canonical path integral and possible gauges}
The canonical Hamiltonian for the gravitational part of the action
\eqref{SDUG} is written as
\begin{equation}
 H_c=\int_{\Sigma_t}d^3x\left( N\cH_T+N^i\cH_i \right) +H_{\cB_t},
\end{equation}
where the Hamiltonian and momentum constraints are defined in
\eqref{cHT} and \eqref{cHi}, and the boundary term in \eqref{HBt}.
The second class constraints are
\begin{equation}\label{scc.DUG}
 p_{\bn}\approx0,\quad
 \cC_\lambda=p_\lambda-\sqrt{h}V_{\bn}\approx0.
\end{equation}
The first class constraints are $\pi_N\approx0,\pi_i\approx0,
p_i\approx0,\cH_i\approx0$ and
\begin{align}
 \cH_T'&=\cH_T-p_{\bn}+\frac{\kappa}{2}\frac{h_{ij}\pi^{ij}}{\sqrt{h}}
 V_{\bn}p_{\bn}\approx0,\nn
 \bar{\cC}'&=\bar{\lambda}-\overline{\left( \frac{p_{\bn}}{\sqrt{h}}
 \right)}\approx0.\label{fcc.DUG}
\end{align}

We introduce gauge fixing conditions as
\begin{align}
 \sigma^0=N-f&\approx0,\quad \sigma^i=N^i-f^i\approx0,\quad
 V^i\approx0,\nn
 \chi^\mu(h_{ij},\pi^{ij})&\approx0,\quad \bar{p}_\lambda\approx0,
 \label{gauge.GRlike}
\end{align}
where $f$ and $f^i$ are fixed functions (or constants), such that
$f>0$, while the conditions $\chi^\mu$ can depend on both $h_{ij}$ and
$\pi^{ij}$, presuming that $\chi^\mu$ depends on $\pi^{ij}$ linearly or
not at all. The four gauge conditions $\chi^\mu$ have to be
independent, so that they fix four components of the variables
$h_{ij},p^{ij}$. Furthermore, it is convenient to require that
\begin{equation}
 \pb{\chi^\mu,\chi^\nu}=0.
\end{equation}

The generator $\bar{\cC}'$ exhibits a nonlocal linear dependence
over the spatial hypersurface, since the spatial integral of the
generator vanishes by definition. The corresponding gauge condition
($\bar{p}_\lambda\approx0$) has a similar nonlocal linear dependence.
Quantization of gauge theories with linearly dependent generators
\cite{Batalin:1984jr} is discussed in \ref{appendix} (see also
\cite{Barvinsky:2010yx}). There we show that the following path integral
is obtained when certain additional gauge conditions are imposed on the
ghost fields associated with the generator $\bar{\cC}'$.

Since the first class constraints have vanishing Poisson brackets with
every constraint except the gauge conditions, we use the Faddeev formula
for the functional determinant of constraints in the path integral.
Furthermore, the determinant of the Poisson bracket between the gauge
conditions and the gauge generators has a block diagonal form, so that
it factorizes. Hence we obtain the integration measure as
\begin{multline}\label{PImeasure.DUG}
\prod_{x^\mu}\cD N\cD\pi_{N}\cD N^{i}\cD\pi_{i}\cD V_{\bn}
\cD p_{\bn} \cD V^{i}\cD p_{i}\cD h_{ij}\cD\pi^{ij}
\cD\lambda_{0}\cD p_{\lambda}^{0}
\cD\bar{\lambda}\cD\bar{p}_{\lambda}\\
\times\delta(p_{\bn})\delta(\cC_{\lambda})
\delta(\pi_{N})\delta(\pi_{i})\delta(\sigma^{\mu})
\delta(p_{i})\delta(V^{i})
\delta(\cH_{\mu}^{\prime})\delta(\chi^{\mu})
\delta(\bar\cC^{\prime})\delta(\bar{p}_{\lambda})\\
\times\sqrt{h}\left|\det\pb{\chi^{\mu},\cH_{\nu}^{\prime}}\right|,
\end{multline}
where we denote $\cH_{\nu}^{\prime}=(\cH_{T}^{\prime},\cH_{i})$.
Integration over the variables $N,\pi_N,N^i,\pi_i,V_{\bn},p_{\bn}$,
$V^i,p_i,\bar\lambda,\bar{p}_\lambda$ can be performed using
the constraint $\delta$-functions. The Hamiltonian and momentum
constraints reduce to \eqref{cHT2} and \eqref{cHi2}, respectively, and
we denote them collectively as $\cH_\nu=(\cH_T,\cH_i)$. In addition, we
write $\delta(\cH_{\mu})$ as an integral over auxiliary fields
$N^\mu=(N,N^i)$, essentially reintroducing the lapse and shift
functions. The auxiliary fields are displaced so that the gauge fixing
functions $(f,f^i)$ are cancelled, $N+f\rightarrow N$ and
$N^i+f^i\rightarrow N^i$. Then the path integral is written as
\begin{multline}\label{ZDUG.first}
 Z_\mathrm{DUG}=\cN^{-1}\int\prod_{x^\mu}\cD N\cD N^i \cD
h_{ij}\cD\pi^{ij}\cD\lambda_{0}
 \cD p_{\lambda}^{0} \delta(\chi^{\mu})
 \left|\det\pb{\chi^{\mu},\cH_{\nu}}\right|\\
 \times \exp\left[ \frac{i}{\hbar}\int dt\left( \int_{\Sigma_t}\left(
 \pi^{ij}\partial_th_{ij} +p_\lambda^0\partial_t\lambda_0
 -N\cH_T-N^i\cH_i \right) -H_{\cB_t}\right) \right],
\end{multline}
where $\cN$ is a normalization factor.
Integration over $p_\lambda^0$ gives a $\delta$-function that imposes
$\partial_t\lambda_0=0$. Therefore we decompose $\lambda_0$ to a
constant component and an average free component $\bar{\lambda}_0$ over
time as
\begin{equation}\label{lambda0dec}
\lambda_0(t)=\frac{2}{\kappa}\Lambda+\bar{\lambda}_0(t),
\end{equation}
where $\int dt\bar{\lambda}_{0}=0$. The integration over
$\bar{\lambda}_{0}$ is performed, which gives
\begin{multline}
 Z_\mathrm{DUG}=\cN^{-1}\int\prod_{x^\mu}\cD N\cD N^i \cD
h_{ij}\cD\pi^{ij}
 \delta(\chi^{\mu}) \left|\det\pb{\chi^{\mu},\cH_{\nu}}\right|\\
 \times \exp\left[ \frac{i}{\hbar}\int dt\left( \int_{\Sigma_t}\left(
 \pi^{ij}\partial_th_{ij} -N\cH_T-N^i\cH_i \right) -H_{\cB_t}\right)
 \right], \label{ZDUG}
\end{multline}
where we have redefined
\begin{equation}\label{cHT3}
 \cH_T=\frac{\kappa}{\sqrt{h}}\pi^{ij}\cG_{ijkl}\pi^{kl}
 -\frac{\sqrt{h}}{\kappa}\sR +\frac{2\sqrt{h}}{\kappa}\Lambda.
\end{equation}
Unlike in \cite{Smolin:2009ti}, the path integral \eqref{ZDUG} does not
include integration over the cosmological constant $\Lambda$, since we
presume that the boundary conditions of the path integral define the
(asymptotic) boundary values of all variables, including the boundary
value of $\lambda$. In particular, the value of $\lambda$ is set to a
constant both on the initial Cauchy surface and on the spatial boundary.
An extension of the path integral with an integral over $\Lambda$ will
be considered in Sect.~\ref{sec3.3.2}.

Then we perform the integration over the momentum $\pi^{ij}$. Since the
measure is at most linear in the momentum, the integration is Gaussian,
and hence the integration can be performed in the standard way (see
e.g. \cite{Shirazi:2013ira}). The integration amounts to expressing the
momentum as
\begin{equation}\label{pi=K}
 \pi^{ij}=\frac{\sqrt{h}}{\kappa}\cG^{ijkl}K_{kl},
\end{equation}
and including the factor $N^{-3}h^{-\frac{1}{2}}$.\footnote{The factor
comes from
$\left|\det\frac{N\kappa}{\sqrt{h}}\cG_{ijkl}\right|^{-\frac{1}{2}}
=2^2\kappa^{-3}N^{-3}h^{-\frac{1}{2}}$, where the de Witt metric
$\cG_{ijkl}$ is regarded as a symmetric 6 by 6 matrix with indices
$(ij)$ and $(kl)$ ranging over the six unique components.}
This result into
\begin{multline}
 Z_\mathrm{DUG}=\cN^{-1}\int\prod_{x^\mu}\cD N\cD N^i \cD h_{ij}
 N^{-3}h^{-\frac{1}{2}} \delta(\chi^{\mu})
 \left|\det\pb{\chi^{\mu},\cH_{\nu}}\right|\\
 \times \exp\left[ \frac{i}{\hbar}\left( \frac{1}{\kappa}\int dt
 \int_{\Sigma_t}N\sqrt{h} \left( K_{ij}\cG^{ijkl}K_{kl}+\sR -2\Lambda
 \right) +S_{\cB} \right) \right].
\end{multline}
Then we express the field differentials as
\begin{equation}
 \cD g_{\mu\nu}=2hN\cD N\cD N^i\cD h_{ij},
\end{equation}
and write $N^{-4}h^{-\frac{3}{2}}=Ng^{00}(-g)^{-\frac{3}{2}}$, and
obtain the path integral as
\begin{equation}\label{ZDUG2}
 Z_\mathrm{DUG}=\cN^{-1}\int\prod_{x^\mu}\cD g_{\mu\nu}
 g^{00}(-g)^{-\frac{3}{2}} N\delta(\chi^{\mu})
 \left|\det\pb{\chi^{\mu},\cH_{\nu}}\right|
 \exp\left( \frac{i}{\hbar}S_\mathrm{EH}[g_{\mu\nu},\Lambda] \right),
\end{equation}
where $S_\mathrm{EH}[g_{\mu\nu},\Lambda]$ is the \EH (EH) action with
an unspecified cosmological constant $\Lambda$,
\begin{equation}
 S_\mathrm{EH}[g_{\mu\nu},\Lambda]=\frac{1}{\kappa}\int_{\cM}d^4x
 \sqrt{-g} \left(R-2\Lambda\right)
 +\frac{2}{\kappa}\oint_{\partial\cM}d^3x\sqrt{|\gamma|}\cK.
\end{equation}
In summary, the difference compared to GR is that the value of the
cosmological constant $\Lambda$ is included in the initial and boundary
conditions, rather than being a coupling constant of the Lagrangian.

The next step is to express the gauge fixing factor of \eqref{ZDUG2} in
a more useful form. For that purpose we consider specific gauge
conditions for the Hamiltonian and momentum constraints. The present
theory has the advantage of enabling the use of the same gauges for the
diffeomorphism symmetry as in GR.

\paragraph{Dirac gauge}
First we consider the Dirac gauge \cite{Dirac:1958jc}:
\begin{equation}\label{gauge.Dirac}
\chi^{0}_{\mathrm{D}}=h_{ij}\pi^{ij}\approx0,\quad
\chi^{i}_{\mathrm{D}}=\partial_{j}\left(h^{ \frac{1}{3}}h^{ij}\right)
\approx0.
\end{equation}
We define an operator $Q^\mu_{\mathrm{D}\,\nu}$ in terms of the gauge
transformation of the gauge conditions \eqref{gauge.Dirac} as
\begin{equation}
 Q^\mu_{\mathrm{D}\,\nu}\xi^\nu
 =\pb{\chi^\mu_{\mathrm{D}},
 \int_{\Sigma_t}\cH_\nu\xi^\nu},\quad
 \xi^\mu=\left( \xi,\xi^i \right).
\end{equation}
Evaluating the Poisson brackets, we obtain the components of the
operator as (up to the Hamiltonian constraint)
\begin{align}
 Q^0_{\mathrm{D}\,0}&=-\frac{2}{\kappa}\sqrt{h}\left( h^{ij}D_iD_j
 -\sR+3\Lambda \right),\label{QD00}\\
 Q^0_{\mathrm{D}\,i}&=-\chi^{0}_{\mathrm{D}}\partial_i
 -\partial_i\chi^{0}_{\mathrm{D}},\label{QD0i}\\
 Q^i_{\mathrm{D}\,0}&=-2h^{\frac{1}{3}} \left( K^{ij}
 -\frac{1}{3}h^{ij}K \right)\partial_j
 -2\partial_j\left[ h^{\frac{1}{3}} \left( K^{ij}
 -\frac{1}{3}h^{ij}K \right) \right],\\
 Q^i_{\mathrm{D}\,j}&=-h^{\frac{1}{3}}\left( \delta^i_{\,j}
 h^{kl}\partial_k\partial_l
 +\frac{1}{3}h^{ik}\partial_k\partial_j \right)
 -\delta^i_{\,j}\chi^{k}_{\mathrm{D}}\partial_k
 +\frac{2}{3}\chi^{i}_{\mathrm{D}}\partial_j
 +\partial_j\chi^{i}_{\mathrm{D}},\label{QDij}
\end{align}
where the momentum $\pi^{ij}$ is written in terms of the metric
\eqref{pi=K} and we denote $K^{ij}=h^{ik}h^{jl}K_{kl}$.
In order to obtain a gauge invariant form for the functional
determinant \cite{Fradkin:1974df}, an extra factor $N$ is included
into the components $Q^0_{\mathrm{D}\,\mu}$.
% , and further the constant
% factor is absorbed into the normalization of the path integral.
Hence we replace \eqref{QD00} with
\begin{equation}
 Q^0_{\mathrm{D}\,0}=-\frac{2}{\kappa}\sqrt{-g}\left(
 h^{ij}D_iD_j-\sR+3\Lambda \right),
\end{equation}
and \eqref{QD0i} with $Q^0_{\mathrm{D}\,i}
=-N\chi^{0}_{\mathrm{D}}\partial_i-N\partial_i\chi^{0}_{\mathrm{D}}$.
For practical applications, the components of the operator
$Q^\mu_{\mathrm{D}\,\nu}$ could be simplified by using the constraints,
in particular the gauge conditions \eqref{gauge.Dirac}, and even
further using the quasiclassical approximation (see
\cite{Fradkin:1974df}).

Finally, the path integral can be written as
\begin{multline}
 Z_\mathrm{DUG}=\cN^{-1}\int\prod_{x^\mu}\cD g_{\mu\nu}\cD\eta_\rho
 \cD c^{*}\cD c\cD c^{*}_i \cD c^i g^{00}(-g)^{-\frac{3}{2}}
 \exp\bigg[ \frac{i}{\hbar}\bigg( S_\mathrm{EH}[g_{\mu\nu},\Lambda]\\
 -\int_{\cM}d^4x\left( \eta_\mu\chi_{\mathrm{D}}^\mu
 +c^{*}Q^0_{\mathrm{D}\,0}c +c^{*}Q^0_{\mathrm{D}\,i}c^i
 +c^{*}_i Q^i_{\mathrm{D}\,0}c +c^{*}_i Q^i_{\mathrm{D}\,j}c^j \right)
 \bigg) \bigg],
\end{multline}
where we have introduced pairs of anti-commuting fields $c,c^{*}$ and
$c^i,c^{*}_i$, commonly referred to as Faddeev-Popov ghosts (and
anti-ghosts), and an auxiliary field $\eta_\mu$ for each gauge
condition $\chi^\mu_{\mathrm{D}}$. Evidently, the full expression for
the action is noncovariant in the Dirac gauge.

\paragraph{Transverse harmonic gauge:}
% Then we consider the covariant transverse harmonic gauge,
\begin{equation}\label{gauge.harmonic}
 \chi^\mu=\partial_\nu\hg^{\mu\nu}\approx0\,;\quad
 \hg^{\mu\nu}=\sqrt{-g}g^{\mu\nu}.
\end{equation}
Transforming to this covariant gauge is achieved via the Faddeev-Popov
trick in the same way as in GR. The operator corresponding to this gauge
is again obtained from the gauge transformation of the gauge conditions
\eqref{gauge.harmonic} as
\begin{equation}
 Q^\mu_{\ \nu}\xi^\nu=\delta_\xi\chi^\mu
 =\partial_\nu\left( \partial_\rho(\hg^{\mu\nu}\xi^\rho)
 -\hg^{\mu\rho}\partial_\rho \xi^\nu
 -\hg^{\rho\nu}\partial_\rho \xi^\mu \right).
\end{equation}
Thanks to the gauge invariant form of the integration measure
\cite{Fradkin:1974df}, the path integral is obtained as
\begin{multline}\label{ZDUGfinal}
 Z_\mathrm{DUG}=\cN^{-1}\int\prod_{x^\mu}\cD g_{\mu\nu}\cD\eta_\rho
 \cD c^{*}_\sigma \cD c^\sigma g^{00}(-g)^{-\frac{3}{2}} \exp\bigg[
 \frac{i}{\hbar}\bigg( S_\mathrm{EH}[g_{\mu\nu},\Lambda] \\
 +\int_{\cM}d^4x\left( -\eta_\mu\chi^\mu
 +\partial_\mu c^{*}_\nu \left( \partial_\rho(\hg^{\mu\nu}c^\rho)
 -\hg^{\mu\rho}\partial_\rho c^\nu
 -\hg^{\rho\nu}\partial_\rho c^\mu \right)
 \right) \bigg) \bigg],
\end{multline}
where $c^\mu$ and $c^{*}_\mu$ are the Faddeev-Popov ghosts.

Matter can be included similarly as in GR. For simplicity we assume that
no extra gauge symmetries or constraints are involved. Finally, we
define the generating functional by including external source
$J^{\mu\nu}$ and $J_\Psi$ for the metric and the matter fields $\Psi$,
respectively,
\begin{multline}\label{ZDUGmatterJ}
 Z_\mathrm{DUG}[J]=\int\prod_{x^\mu}\cD g_{\mu\nu}\cD\eta_\rho
 \cD c^{*}_\sigma \cD c^\sigma \cD\Psi g^{00}(-g)^{-\frac{3}{2}}
  \exp\bigg[
 \frac{i}{\hbar}\bigg( S_\mathrm{EH}[g_{\mu\nu},\Lambda]
 +S_\mathrm{m}[g_{\mu\nu},\Psi] \\
 +\int_{\cM}d^4x\left( -\eta_\mu\chi^\mu
 +\partial_\mu c^{*}_\nu \left( \partial_\rho(\hg^{\mu\nu}c^\rho)
 -\hg^{\mu\rho}\partial_\rho c^\nu
 -\hg^{\rho\nu}\partial_\rho c^\mu \right) \right.\\
 +\left. g_{\mu\nu}J^{\mu\nu} +\Psi J_\Psi \right) \bigg) \bigg].
\end{multline}

We have shown that the path integral for the fully diffeomorphism
invariant unimodular gravity \eqref{SDUG} has the same form as the path
integral for GR with a cosmological constant. The crucial difference
from GR is that the value of the cosmological constant is set as a part
of the boundary conditions for the path integral.

The quantum effective action for the DUG theory can be defined in the
exact same way as for GR, since there are no extra conditions on the
metric and the path integrals have the same form.

\subsubsection{Ng and van Dam form of the path integral}\label{sec3.3.2}
The path integral for unimodular gravity can be extended by including an
integration over the cosmological constant $\Lambda$. Then the path
integral takes the following form:
\begin{equation}\label{ZNvD}
 Z_{\mathrm{NvD}}=\int d\mu(\Lambda) Z_\mathrm{DUG}(\Lambda),
\end{equation}
where $d\mu(\Lambda)$ is an integration measure for $\Lambda$, and the
path integral for unimodular gravity, $Z_\mathrm{DUG}(\Lambda)$, is
given in \eqref{ZDUGfinal} with boundary conditions chosen to be
consistent with a given value $\Lambda$ of the cosmological constant.
This form of the path integral for unimodular gravity was originally
proposed in \cite{Ng:1990rw,Ng:1990xz}. It was also later derived from a
canonical path integral \cite{Smolin:2009ti}, although some manipulation
of variables was required, and the canonical measure was assumed to
include an integral over $\Lambda$. In \cite{Smolin:2009ti}, the
integral over $\lambda_0$ was assumed to include integration over both
$\Lambda$ and $\bar\lambda_0$ due to the decomposition
\eqref{lambda0dec}. Here we show that the path integral \eqref{ZNvD}
follows straightforwardly from the canonical path integral of the action
\eqref{SDUG}, when we consider the vacuum state of the universe to be a
superposition of the states corresponding to different values of
$\Lambda$ \cite{Weinberg:1988cp}.

We emphasize that \eqref{ZNvD} is a quite different path integral
compared to the one we derived above \eqref{ZDUGfinal}. In the path
integral \eqref{ZDUGfinal}, the value of the cosmological constant is
set as a part of the physical boundary conditions, which (together with
a semiclassical matter distribution) define the vacuum state of the
system. Including an additional integration over $\Lambda$ means that we
are integrating over different boundary conditions, i.e., vacuums. Below
we attempt to justify the integration of $\Lambda$ properly.

Let $\ket{\Lambda}$ denote the vacuum state of the universe that is
consistent with a given value of the cosmological constant $\Lambda$
and with other relevant boundary conditions. The path integral that we
have obtained for the fully diffeomorphism invariant unimodular gravity
\eqref{ZDUGfinal} represents the vacuum transition amplitude
\begin{equation}\label{vevLambda}
 \braket{\Lambda|\Lambda}=Z_\mathrm{DUG}(\Lambda).
\end{equation}
We assume that transitions between vacuums are prohibited if the vacuums
correspond to different values of $\Lambda$, i.e., the states
$\ket{\Lambda}$ are assumed to be orthogonal,
\begin{equation}\label{orthoLambda}
 \braket{\Lambda|\Lambda'}=0\quad\mathrm{if}\ \Lambda\neq\Lambda'.
\end{equation}
Furthermore we assume that the states are nondegenerate, i.e., there
exists one state $\ket{\Lambda}$ for each value of $\Lambda$.
The vacuum state of the universe is written as a superposition of the
states corresponding to different values of $\Lambda$ as
\begin{equation}\label{vacOmega}
 \ket{\Omega}=\int d\Lambda\,\omega(\Lambda)\ket{\Lambda}.
\end{equation}
Choosing the weight function $\omega(\Lambda)$ defines which states
$\ket{\Lambda}$ are included in the superposition. Now the vacuum
transition amplitude is obtained in the form \eqref{ZNvD} as
\begin{equation}\label{vevOmega}
 \braket{\Omega|\Omega}=\int d\Lambda|\omega(\Lambda)|^2
 \braket{\Lambda|\Lambda}
 =\int d\mu(\Lambda)Z_\mathrm{DUG}(\Lambda),
\end{equation}
where the measure $d\mu(\Lambda)$ is defined by the weight function as
\begin{equation}
 d\mu(\Lambda)=|\omega(\Lambda)|^2 d\Lambda.
\end{equation}

Using the semiclassical approximation and then the stationary phase
approximation, it was argued in \cite{Ng:1990rw} that the path integral
\eqref{ZNvD} for pure gravity is dominated by solutions whose
cosmological constant $\Lambda=0$. In the presence of matter
\eqref{ZDUGmatterJ}, the same argument was used in \cite{Smolin:2009ti}
to obtain that the path integral \eqref{ZNvD} is dominated by the
solutions of the Einstein equation whose cosmological constant is
approximately
\begin{equation}\label{Lambda.dom}
 \Lambda=2\pi G\frac{\int_{\cM}\sqrt{-g}\rho}{\int_{\cM}\sqrt{-g}},
\end{equation}
where $\rho$ is the energy density of a perfect fluid. This result was
argued to imply that \eqref{Lambda.dom} is the most likely value of the
cosmological constant. It is intriguing that using the present day
energy density as an estimate for the average density
\cite{Smolin:2009ti}, one obtains a result that is surprisingly close to
the observed value of $\Lambda$ (the observed $\Lambda/G$ being about
three times the present average energy density).

The result \eqref{Lambda.dom} is based on a hidden assumption that the
given value of $\Lambda$ is included in the vacuum state
\eqref{vacOmega}. It was assumed that all states $\ket{\Lambda}$ are
weighted equally, $|\omega(\Lambda)|^2=\mathrm{constant}$. This
corresponds to a total lack of physical boundary conditions regarding
$\Lambda$, and then using the path integral for finding the most likely
value of $\Lambda$. This is an interesting argument, but speculative
and conceptually problematic. We indeed need information on the boundary
conditions in order to estimate the average value of $\rho$ over
spacetime. Even if we accept \eqref{Lambda.dom} as a valid estimate for
the value of $\Lambda$ in our universe, estimating the average of
matter energy density over the whole spacetime is challenging, to say
the least.

\section{Quantization of the unimodular gravity with a fixed metric
determinant}
\label{sec4}
\subsection{ADM decomposition of the action}\label{sec4.1}
The gravitational part of the action \eqref{SUG} is written in ADM form
as
\begin{multline}\label{SUG.ADM}
 S_\mathrm{UG}[N,N^i,h_{ij},\lambda,\Psi]=\int dt\int_{\Sigma_t}d^3x
 \left[ \frac{N\sqrt{h}}{\kappa}\left( K_{ij}\cG^{ijkl}K_{kl}+\sR
 \right) -\lambda\left( N\sqrt{h}-\epsilon_0 \right) \right]\\
 +S_{\cB} +S_\mathrm{m}[g_{\mu\nu},\Psi].
\end{multline}
A Hamiltonian formulation of an action of this form has been considered
in \cite{Kluson:2014esa}, and our following analysis is similar in
several ways.

\subsection{Hamiltonian analysis}\label{sec4.2}
The momenta conjugate to $N$, $N^i$, and $\lambda$ are the primary
constraints:
\begin{equation}
 \pi_N\approx0,\quad \pi_i\approx0,\quad p_\lambda\approx0.
\end{equation}
The Hamiltonian is obtained as
\begin{equation}\label{HUG}
 H=\int_{\Sigma_t}d^3x\left( N\cH_T+N^i\cH_i -\epsilon_0\lambda
 +v_N\pi_N+v_N^i\pi_i +v_\lambda p_\lambda \right) +H_{\cB_t},
\end{equation}
where the super-Hamiltonian is defined as
\begin{equation}\label{cHTu}
  \cH_T=\frac{\kappa}{\sqrt{h}}\pi^{ij}\cG_{ijkl}\pi^{kl}
 -\frac{\sqrt{h}}{\kappa}\sR +\sqrt{h}\lambda,
\end{equation}
the supermomentum is defined as
\begin{equation}\label{cHiu}
 \cH_i=-2h_{ij}D_k\pi^{jk},
\end{equation}
$v_N,v_N^i,v_\lambda$ are Lagrange multipliers, and $\epsilon_0$ is the
fixed scalar density.

Preservation of the primary constraints implies the secondary
constraints:
\begin{equation}\label{scu}
 \cH_T\approx0,\quad \cH_i\approx0,\quad
\cU=N\sqrt{h}-\epsilon_0\approx0.
\end{equation}
The momentum constraint \eqref{cHiu} can again be extended with
terms that are proportional to the primary constraints $\pi_N$ and
$p_\lambda$,
\begin{equation}\label{cHiu2}
 \cH_i=-2h_{ij}D_k\pi^{jk} +\partial_iN\pi_N
 +\partial_i\lambda p_\lambda,
\end{equation}
since then it will generate spatial diffeomorphisms on $\Sigma_t$ for
all the variables that are involved in the secondary constraints
\eqref{scu}. The smeared Hamiltonian and momentum constraints
\eqref{HMC} satisfy the following Poisson brackets:
\begin{align}
 \pb{\cH_T[\xi],\cH_T[\eta]}&=\int_{\Sigma_t}d^3x
 (\xi\partial_i\eta -\eta\partial_i\xi)h^{ij}(\cH_j -\partial_jN\pi_N
 -\partial_j\lambda p_\lambda),\nn
 \pb{\Phi[\chi^i],\cH_T[\xi]}&=\cH_T[\chi^i\partial_i\xi],\nn
 \pb{\Phi[\chi^i],\Phi[\psi^j]}&=\Phi[\chi^j\partial_j\psi^i
 -\psi^j\partial_j\chi^i]. \label{scalgebra2}
\end{align}
The Hamiltonian and momentum constraints have nonvanishing Poisson
brackets with $\cU$:
\begin{align}
 \pb{\cU,\cH_T[\xi]}&=-\frac{\kappa}{2} N\xi h_{ij}\pi^{ij}
 \approx -\frac{\kappa}{2}
\epsilon_0\xi\frac{h_{ij}\pi^{ij}}{\sqrt{h}},\\
  \pb{\cU,\Phi[\chi^i]}&=\chi^i\partial_i(N\sqrt{h})
  +\partial_i\chi^iN\sqrt{h} %\approx \partial_i\chi^i N\sqrt{h}.
  \approx \epsilon_0\partial_i\chi^i.
\end{align}
Hence the preservation of $\cU$,
\begin{equation}
 \partial_t\cU=\pb{\cU,H}\approx -N\frac{\kappa \epsilon_0}{2\sqrt{h}}
 h_{ij}\pi^{ij} +\epsilon_0\partial_iN^i
 +\sqrt{h}v_N\approx0,
\end{equation}
is ensured by fixing the Lagrange multiplier $v_N$ as
\begin{equation}
 v_N=w_N\equiv N\frac{\kappa \epsilon_0}{2h}h_{ij}\pi^{ij}
 -\frac{\epsilon_0}{\sqrt{h}}\partial_iN^i.
\end{equation}
The preservation of $\cH_T$,
\begin{equation}
 \partial_t\cH_T=\pb{\cH_T,H}\approx\sqrt{h}v_\lambda\approx0,
\end{equation}
fixes the Lagrange multiplier $v_\lambda$ as
\begin{equation}\label{vlambda2}
 v_\lambda=0.
\end{equation}
The preservation of $\cH_i\approx0$,
\begin{equation}
 \partial_t\cH_i=\pb{\cH_i,H}\approx
\epsilon_0\partial_i\lambda\approx0,
\end{equation}
requires the introduction of the secondary constraint \eqref{cCi}, which
was also present in the generally covariant formulation. The constraint
\eqref{cCi} is preserved in time since the Lagrange multiplier of the
primary constraint $p_\lambda\approx0$ has been fixed to zero
\eqref{vlambda2}. We do not need any further constraints, but we still
need to analyze and classify the existing constraints properly.

We again decompose the variables $\lambda,p_\lambda$ as in
\eqref{lambdadec} and replace the constraint \eqref{cCi} with
\eqref{barlambda}. The second class constraints $\bar\lambda\approx0$,
$\bar{p}_\lambda\approx0$ can be used to eliminate
the average free variables $\bar\lambda,\bar{p}_\lambda$. Since the
Hamiltonian constraint $\cH_T$ contains the remaining zero mode
$\lambda_0$, and the zero mode $p_\lambda^0$ of the primary
constraint $p_\lambda\approx0$ remains, we should also decompose
$\cH_T$ as
\begin{equation}\label{cHTdec}
 \cH_T=\frac{\sqrt{h}}{\int_{\Sigma_t}d^3x\sqrt{h}}\cH_0
 +\bcH_T,\quad \cH_0=\int_{\Sigma_t}d^3x\cH_T,\quad
 \int_{\Sigma_t}d^3x\bcH_T=0,
\end{equation}
where the zero mode and the average free component are respectively
defined as
\begin{equation}\label{cH0}
 \cH_0=\int_{\Sigma_t}d^3x\left(
 \frac{\kappa} {\sqrt{h}}\pi^{ij}\cG_{ijkl}\pi^{kl}
 -\frac{\sqrt{h}}{\kappa}\sR \right)
 +\lambda_0\int_{\Sigma_t}d^3x\sqrt{h}\approx0
\end{equation}
and
\begin{equation}\label{bcH_T}
\begin{split}
 \bcH_T&=\overline{\frac{\kappa}{\sqrt{h}}\pi^{ij}\cG_{ijkl}\pi^{kl}}
 -\overline{\frac{\sqrt{h}}{\kappa}\sR}\\
 &=\frac{\kappa} {\sqrt{h}}\pi^{ij}\cG_{ijkl}\pi^{kl}
 -\frac{\sqrt{h}}{\kappa}\sR
 -\frac{\sqrt{h}}{\int_{\Sigma_t}d^3x\sqrt{h}}
 \int_{\Sigma_t}d^3x\left(
 \frac{\kappa} {\sqrt{h}}\pi^{ij}\cG_{ijkl}\pi^{kl}
 -\frac{\sqrt{h}}{\kappa}\sR \right)
 \approx0.
% \\ \int_{\Sigma_t}d^3x\bcH_T&=0.
\end{split}
\end{equation}
Note that \eqref{bcH_T} does not involve the cosmological variable
$\lambda_0$.
The components $\cH_0$ and $\bcH_T$ satisfy the Poisson brackets
\begin{align}
 \pb{\cH_0,\cH_0}&=0,\nn
 \pb{\cH_0,\bcH_T[\bar\xi]}&=\int_{\Sigma_t}d^3x\partial_i\bar\xi
 h^{ij}(\cH_j -\partial_jN\pi_N -\partial_j\lambda p_\lambda),\nn
 \pb{\bcH_T[\bar\xi],\bcH_T[\bar\eta]}&=
 \int_{\Sigma_t}d^3x (\bar\xi\partial_i\bar\eta
 -\bar\eta\partial_i\bar\xi)h^{ij}(\cH_j -\partial_jN\pi_N
 -\partial_j\lambda p_\lambda),
\end{align}
where the smeared constraint $\bcH_T[\bar\xi]$ is defined so that
\begin{equation}
 \cH_T[\xi]=\xi_0\cH_0+\bcH_T[\bar\xi],\quad
 \bcH_T[\bar\xi]=\int_{\Sigma_t}d^3x\bar\xi\bcH_T,
\end{equation}
and where $\xi$ is decomposed as any scalar, $\xi=\xi_0+\bar{\xi}$, and
\begin{equation}
 \xi_0=\frac{1}{\int_{\Sigma_t}d^3x\sqrt{h}}
 \int_{\Sigma_t}d^3x\sqrt{h}\xi,\quad
 \int_{\Sigma_t}d^3x\sqrt{h}\bar{\xi}=0.
\end{equation}

In the Hamiltonian \eqref{HUG}, we obtain
\begin{equation}
 \int_{\Sigma_t}d^3x\left( N\cH_T +N^i\cH_i +w_N\pi_N \right)
 =\int_{\Sigma_t}d^3x\left( N\cH'_T +N^i\cH'_i \right),
\end{equation}
where we have extended the Hamiltonian and momentum constraints as
\begin{align}
 \cH'_T&=\cH_T+\frac{\kappa}{2} \epsilon_0
\frac{h_{ij}\pi^{ij}}{\sqrt{h}}
 \frac{\pi_N}{\sqrt{h}}\approx0,\nn
 \cH'_i& =\cH_i+\epsilon_0\partial_i\left(\frac{\pi_N}{\sqrt{h}}\right)
 \approx0,
\end{align}
and furthermore
\begin{equation}
 \int_{\Sigma_t}d^3x N\cH'_T=N_0\cH'_0
 +\int_{\Sigma_t}d^3x\bar{N}\bcH'_T,
\end{equation}
where the zero mode and average free component of $\cH'_T$ are defined
in the same way as for $\cH_T$ in \eqref{cHTdec}--\eqref{bcH_T}.
The first class constraints are the average free Hamiltonian constraint
$\bcH'_T$ and the constraints $\cH'_i$, $\pi_i$. These constraints are
associated with the invariance of the action \eqref{SUG} under the
metric determinant-preserving diffeomorphism
\eqref{unimodvar}--\eqref{divxi}. Since the lapse $N$ is not an
unspecified multiplier in the Hamiltonian, we should add the term
$\bar{v}_T\bcH'_T$ into the Hamiltonian density, where $\bar{v}_T$ is an
unspecified Lagrange multiplier. The Hamiltonian \eqref{HUG} is
rewritten as
\begin{equation}\label{HUG2}
 H=\int_{\Sigma_t}d^3x\left( N\cH'_T -\epsilon_0\lambda_0
+\bar{v}_T\bcH'_T +N^i\cH'_i +v_N^i\pi_i \right) +H_{\cB_t}.
\end{equation}

The local constraints $\pi_N\approx0,\cU\approx0$ and the zero mode
constraints $p_\lambda^0\approx0,\cH_0\approx0$ are the second class
constraints. The second class constraints $\pi_N\approx0,\cU\approx0$
can be used to eliminate the variables $N,\pi_N$ as
\begin{equation}\label{N.fixed}
 N=\frac{\epsilon_0}{\sqrt{h}},\quad \pi_N=0.
\end{equation}
The zero mode constraints $p_\lambda^0\approx0,\cH_0\approx0$ can be
used to eliminate the variables $\lambda_0,p_\lambda^0$ as
\begin{equation}\label{lambda0.fixed}
 \lambda_0=-\frac{1}{\int_{\Sigma_t}d^3x\sqrt{h}}
 \int_{\Sigma_t}d^3x\left( \frac{\kappa}{\sqrt{h}}
 \pi^{ij}\cG_{ijkl}\pi^{kl} -\frac{\sqrt{h}}{\kappa}\sR \right),\quad
 p_\lambda^0=0.
\end{equation}
The Dirac bracket that corresponds to the second class constraints
($\pi_N\approx0,\cU\approx0,p_\lambda^0\approx0,\cH_0\approx0$)
can be shown to be equal to Poisson bracket for all the remaining
variables.

When the second class constraints are set to zero strongly and the
auxiliary variables are eliminated as \eqref{N.fixed} and
\eqref{lambda0.fixed}, we obtain the Hamiltonian as
\begin{equation}\label{HUG3}
\begin{split}
 H&=\frac{\int_{\Sigma_t}d^3x\epsilon_0}{\int_{\Sigma_t}d^3x\sqrt{h}}
 \int_{\Sigma_t}d^3x\left( \frac{\kappa}{\sqrt{h}}
 \pi^{ij}\cG_{ijkl}\pi^{kl} -\frac{\sqrt{h}}{\kappa}\sR \right)\\
 &\quad+\int_{\Sigma_t}d^3x\left[ \left(\frac{\epsilon_0}{\sqrt{h}}
 +\bar{v}_T\right)\bcH_T +N^i\cH_i +v_N^i\pi_i \right] +H_{\cB_t}.
\end{split}
\end{equation}

\subsection{Path integral}\label{sec4.3}
The canonical Hamiltonian for the action \eqref{SUG} is written as
\begin{equation}
 H_c=\int_{\Sigma_t}d^3x\left( N\cH_T+N^i\cH_i-\epsilon_0\lambda
 \right) +H_{\cB_t},
\end{equation}
where the Hamiltonian and momentum constraints are defined in
\eqref{cHTu} and \eqref{cHiu}.
The second class constraints are
\begin{align}\label{scc.UG}
 \cU=N\sqrt{h}-\epsilon_0&\approx0,\quad \pi_N\approx0,\nn
 \bar{\lambda}&\approx0,\quad \bar{p}_\lambda\approx0,\nn
 \cH_0&\approx0,\quad p_\lambda^0\approx0.
\end{align}
The first class constraints are $\pi_i\approx0$ and
\begin{align}
 \bcH'_T&=\bcH_T+\overline{\frac{\kappa}{2}\epsilon_0
 \frac{h_{ij}\pi^{ij}}{\sqrt{h}} \frac{\pi_N}{\sqrt{h}}}\approx0,\nn
 \cH'_i& =\cH_i+\epsilon_0\partial_i\left(\frac{\pi_N}{\sqrt{h}}\right)
 \approx0.\label{fcc.UG}
\end{align}
We denote the latter two constraints collectively as
$\tcH'_\mu=(\bcH'_T,\cH'_i)$.

The gauge fixing condition for $N^i$ is defined as in
\eqref{gauge.GRlike}, but there is no gauge condition for $N$ due to
the first pair of second class constraints in \eqref{scc.UG}.
The gauge conditions read as
\begin{equation}\label{gauge.UG}
 \sigma^i=N^i-f^i\approx0,\quad
 \tilde\chi^\mu[h_{ij},\pi^{ij}]\approx0,
\end{equation}
where one of the conditions $\tilde\chi^\mu$ has to be average free, so
that the number of gauge conditions matches the number of generators
exactly. We choose it to be the zero-component, since the zero mode of
the super-Hamiltonian is a second class constraint, and hence we
denote $\tilde\chi^\mu=(\bar\chi^0,\chi^i)$.

The generator $\bcH'_T$ and the gauge condition $\bar\chi^0$ both
suffer from a nonlocal linear dependence over the spatial hypersurface,
since their spatial integrals vanish by definition. The proper treatment
of linearly dependent generators \cite{Batalin:1984jr} is discussed in
Appendix~\ref{appendix}.

The canonical integration measure for the path integral is written as
\begin{multline}
\prod_{x^\mu}\cD N\cD\pi_{N}\cD N^{i}\cD\pi_{i}\cD h_{ij}\cD\pi^{ij}
\cD\lambda_{0}\cD p_{\lambda}^{0}
\cD\bar{\lambda}\cD\bar{p}_{\lambda}
\delta(\cU)\delta(\pi_{N})
\delta(\bar\lambda)\delta(\bar{p}_{\lambda})\\
\times\delta(\pi_{i})\delta(\sigma^{j})
\delta(\cH_{0})\delta(p_{\lambda}^{0})
\delta(\tcH_{\mu}^{\prime})\delta(\tilde\chi^{\mu})
\left( \sqrt{h}\int_{\Sigma_t}\sqrt{h} \right)
\left|\det\pb{\tilde\chi^{\mu},\tcH_{\nu}^{\prime}}\right|.
\end{multline}
The initial and boundary conditions on the cosmological variable are
similar to DUG, i.e., the value of $\lambda$ is set to a constant on
the initial Cauchy surface and on the spatial boundary.
When integration over the variables $N,\pi_N$, $N^i$, $\pi_i$,
$\bar\lambda$, $\bar{p}_\lambda$ and $p_\lambda^0$ is performed using
the constraints, we obtain
\begin{multline}\label{ZUG.first}
 Z_\mathrm{UG}=\cN^{-1}\int\prod_{x^\mu}\cD h_{ij}\cD\pi^{ij}
 \cD\lambda_{0} \,\delta(\cH_{0})\delta(\tcH_{\mu})
 \delta(\tilde\chi^{\mu}) \left( \int_{\Sigma_t}\sqrt{h} \right)
 \left|\det\pb{\tilde\chi^{\mu},\tcH_{\nu}}\right|\\
 \times\exp\left[ \frac{i}{\hbar}\int dt\left( \int_{\Sigma_t}\left(
 \pi^{ij}\partial_th_{ij} -\frac{\epsilon_0}{\sqrt{h}}\cH_T-f^i\cH_i
 +\epsilon_0\lambda_0 \right) -H_{\cB_t} \right) \right],
\end{multline}
where we denote $\tcH_\nu=(\bcH_T,\cH_i)$ and the Hamiltonian
constraint is given in \eqref{cHT2}. Expressing the $\delta$-functions
$\delta(\cH_{0})$ and $\delta(\tcH_{\mu})$ in terms of integrals over
auxiliary variables $N=(N_0,\bar{N})$ and $N^i$, and shifting the
variables as $N\rightarrow N-\frac{\epsilon_0}{\sqrt{h}}$ and
$N^i\rightarrow
N^i-f^i$,\footnote{Note that the change of variable $N\rightarrow
N'=N+\frac{\epsilon_0}{\sqrt{h}}$ has a unit Jacobian, despite the
fact that the transformation involves $\sqrt{h}$.} we obtain
\begin{multline}
 Z_\mathrm{UG}=\cN^{-1}\int\prod_{x^\mu}\cD N\cD N^i \cD h_{ij}
 \cD\pi^{ij} \cD\lambda_{0} \left( \int_{\Sigma_t}\sqrt{h} \right)
 \delta(\tilde\chi^{\mu})
 \left|\det\pb{\tilde\chi^{\mu},\tcH_{\nu}}\right|\\
 \times\exp\left[ \frac{i}{\hbar}\int dt\left( \int_{\Sigma_t}\left(
 \pi^{ij}\partial_th_{ij} -N\cH_T-N^i\cH_i +\epsilon_0\lambda_0
 \right) -H_{\cB_t} \right) \right].
\end{multline}
Integration over the momentum $\pi^{ij}$ gives
\begin{multline}
 Z_\mathrm{UG}=\cN^{-1}\int\prod_{x^\mu}\cD g_{\mu\nu}\cD\lambda_{0}
 g^{00}(-g)^{-\frac{3}{2}} \left( N\int_{\Sigma_t}\sqrt{h} \right)
 \delta(\tilde\chi^{\mu})
 \left|\det\pb{\tilde\chi^{\mu},\tcH_{\nu}}\right|\\
 \times \exp\left[ \frac{i}{\hbar}\left( S_\mathrm{EH}[g_{\mu\nu}]
 -\int dt\int_{\Sigma_t}\lambda_0\left( \sqrt{-g}-\epsilon_0 \right)
 \right) \right],
\end{multline}
where $S_\mathrm{EH}[g_{\mu\nu}]$ is the \EH action without a
cosmological constant.
Since the zero mode $\lambda_0$ depends only on time, integration over
this variable gives a $\delta$-function that imposes the unimodular
condition \eqref{unimodcond} to hold on each slice $\Sigma_t$ of
spacetime in average,
\begin{multline}\label{ZUG}
 Z_\mathrm{UG}=\cN^{-1}\int\prod_{x^\mu}\cD g_{\mu\nu}
 g^{00}(-g)^{-\frac{3}{2}}
 \delta\left( \frac{\int_{\Sigma_t}\left( \sqrt{-g}-\epsilon_0 \right)}
 {N\int_{\Sigma_t}\sqrt{h}} \right) \delta(\tilde\chi^{\mu})
 \left|\det\pb{\tilde\chi^{\mu},\tcH_{\nu}}\right|\\
 \times \exp\left(\frac{i}{\hbar}S_\mathrm{EH}[g_{\mu\nu}] \right).
\end{multline}
The integrated unimodular condition in the above path integral,
\begin{equation}\label{unimodcond.av}
 \int_{\Sigma_t}\left( \sqrt{-g}-\epsilon_0 \right)=0,
\end{equation}
does not constrain local deviations from the unimodular condition
\eqref{unimodcond} as long as the average value of $\sqrt{-g}$
over $\Sigma_t$ remains fixed to that of $\epsilon_0$.
This is a quite surprising result, since we expected to see the
unimodular condition to be imposed locally, like in the path integral
for the HT action \cite{Smolin:2009ti}. On the other hand, it makes some
sense that quantum fluctuations around the classical field equation
\eqref{unimodcond} are permitted.
The physical purpose of the condition \eqref{unimodcond.av} is to
ensure that the number of physical degrees of freedom in the path
integral \eqref{ZUG} matches that of DUG and GR, since together the
gauge conditions $\tilde\chi^{\mu}$ and the condition
\eqref{unimodcond.av} impose four conditions per point in space.

In this theory, the quantum effective action is a function of the
perturbative gravitational field $f_{\mu\nu}$ which satisfies an
integrated condition. Namely the trace of the perturbative field must
have zero integral over $\Sigma_t$ at all times,
\begin{equation}\label{trf.av}
 \int_{\Sigma_t}f^\mu_{\ \mu}=0.
\end{equation}
In other words, the quantum effective action is built in the same way
as in the HT theory \cite{Smolin:2009ti} except that the condition on
the perturbative gravitational field ($f^\mu_{\ \mu}=0$) is replaced
with the integrated condition \eqref{trf.av}. The gravitational field is
further constrained by the gauge conditions $\tilde\chi^\mu$.

\subsubsection*{Counting of physical degrees of freedom}
In both cases, DUG and UG, Dirac's counting of physical degrees of
freedom gives the same result: two propagating modes plus one
zero/single mode. In DUG, the extra zero mode is the cosmological
variable $\lambda$, which is a constant spatially and does not evolve.
Thus the extra zero mode is not a true physical degrees of freedom. It
is just a cosmological constant. Hence the physical degrees of freedom
are the same as in GR.

In the UG theory with fixed metric determinant, the canonical structure
is partially different from DUG. In particular the integral of the
Hamiltonian constraint is a second class constraint, and hence the
gauge/coordinate conditions must contain one zero mode less than in DUG
and GR. This may appear to imply that the extra zero mode would be a
physical degree of freedom, but our analysis shows otherwise. This is
evident in the path integral \eqref{ZUG}, where the single
$\delta$-function eliminates one zero mode by imposing the integrated
unimodular condition \eqref{unimodcond.av}. In other words it acts like
an extra gauge/coordinate condition, so that the total number of
conditions matches DUG and GR. Thus the actual number of physical
degrees of freedom in UG is the same as in DUG and GR.

\paragraph{Gauge fixing example}
We can choose the gauge conditions, for example, as
\begin{align}\label{gauge.uh}
\bar\chi^{0}_{\mathrm{U}}&=\overline{\sqrt{h}\left( \ln h-\Phi
\right)}=\sqrt{h}\left( \ln h-\Phi -\frac{1}{\int_{\Sigma_t}\sqrt{h}}
\int_{\Sigma_t}\sqrt{h}\left( \ln h-\Phi \right) \right)\approx0,\nn
\chi^{i}_{\mathrm{U}}&=\partial_{j}\left(\sqrt{h}h^{ij}\right)
\approx0,
\end{align}
where $\Phi$ is a fixed function, and we denote the conditions
collectively as $\tilde\chi^{\mu}_{\mathrm{U}}=
(\bar\chi^{0}_{\mathrm{U}}, \chi^{i}_{\mathrm{U}})$.
The first gauge condition $\bar\chi^{0}_{\mathrm{U}}$ fixes the average
free component of $\ln h$. That is the average free component of the
first condition of the Faddeev-Popov gauge \cite{Faddeev:1973zb}.
The gauge conditions $\chi^{i}_{\mathrm{U}}$ are the harmonic conditions
on the spatial hypersurface.

We define an operator $Q^\mu_{\mathrm{U}\,\nu}$ in terms of the gauge
transformation of the gauge conditions \eqref{gauge.uh} as
\begin{equation}
 Q^\mu_{\mathrm{U}\,\nu}\tilde\xi^\nu
 =\pb{\tilde\chi^\mu_{\mathrm{U}},
 \int_{\Sigma_t}\tcH_\nu\tilde\xi^\nu},\quad
 \tilde\xi^\nu=\left(\bar\xi,\xi^i\right).
\end{equation}
We obtain the components of the operator as
\begin{align}
 Q^0_{\mathrm{U}\,0}&=2\sqrt{h}K+\bar\chi^{0}_{\mathrm{U}}K,
 \label{QU00}\\
 Q^0_{\mathrm{U}\,i}&=2\sqrt{h}D_i+\bar\chi^{0}_{\mathrm{U}}D_i,
 \label{QU0i}\\
 Q^i_{\mathrm{U}\,0}&=-2\sqrt{h}\left( K^{ij}-\frac{1}{2}h^{ij}K
 \right)\partial_j
 -2\partial_j\left[ \sqrt{h}\left( K^{ij} -\frac{1}{2}\sqrt{h}h^{ij}K
 \right) \right] \label{QUi0},\\
 Q^i_{\mathrm{U}\,j}&=-\sqrt{h}\delta^i_{\,j}h^{kl}\partial_k\partial_l
 -\delta^i_{\,j}\chi^{k}_{\mathrm{U}}\partial_k
 +\chi^{i}_{\mathrm{U}}\partial_j
 +\partial_j\chi^{i}_{\mathrm{U}}.\label{QUij}
\end{align}
The components \eqref{QU00}--\eqref{QUij} of the operator could be
simplified by using the constraints and in particular the gauge
conditions. Finally, the path integral is written as
\begin{multline}\label{ZUG.Ug}
 Z_\mathrm{UG}=\cN^{-1}\int\prod_{x^\mu}\cD g_{\mu\nu}\cD\bar\eta
 \cD\eta_i \cD\bar{c}^{*}\cD\bar{c}\cD c^{*}_i \cD c^i
 g^{00}(-g)^{-\frac{3}{2}} \delta\left(
 \frac{\int_{\Sigma_t}\left( \sqrt{-g} -\epsilon_0 \right)}
 {N\int_{\Sigma_t}\sqrt{h}} \right)\\
 \exp\bigg[ \frac{i}{\hbar}\bigg( S_\mathrm{EH}[g_{\mu\nu}]
 -\int_{\cM}d^4x\Big( \bar\eta\bar\chi_{\mathrm{U}}^0
 +\eta_i\chi_{\mathrm{U}}^i +\bar{c}^{*}Q^0_{\mathrm{U}\,0}\bar{c}
 +\bar{c}^{*}Q^0_{\mathrm{U}\,i}c^i \\
 +c^{*}_i Q^i_{\mathrm{U}\,0}\bar{c}
 +c^{*}_i Q^i_{\mathrm{U}\,j}c^j \Big) \bigg) \bigg],
\end{multline}
where we have introduced pairs of anti-commuting ghosts $\bar{c},
\bar{c}^{*}$ and $c^i,c^{*}_j$, and auxiliary fields $\bar\eta,\eta_i$
for each gauge condition. The fields $\bar{c}$, $\bar{c}^{*}$,
$\bar\eta$ have vanishing average over space, since they are
associated with the generator $\bcH_T$ and the gauge condition
$\bar\chi_{\mathrm{U}}^0$. Including matter fields and
defining the generating functional can be done similarly as in
\eqref{ZDUGmatterJ}. Evidently, the above expression for the path
integral is not covariant. The presence of integration over space in
both the averaged unimodular condition and the definition of average
free fields renders the expression noncovariant.

It indeed appears to be impossible to cast the path integral \eqref{ZUG}
into a fully covariant form. The underlying reason is the fact that the
zero mode of the super-Hamiltonian is a second class constraint, and
hence one of the gauge conditions must be average free over space. In
order to achieve a covariant description, we have to enlarge the gauge
symmetry so that the total super-Hamiltonian becomes a gauge generator.
This was achieved in Sect.~\ref{sec3}, where a generally covariant
form of unimodular gravity is considered.

\section{The canonical relation of the two theories}\label{sec5}
In the case with a fixed metric determinant, it is crucial to notice
that the Hamiltonian \eqref{HUG} is not a constraint, since it contains
the term $-\int_{\Sigma_t}d^3x\epsilon_0\lambda$. Therefore the bulk
part of the Hamiltonian does not vanish on the constraint surface. This
is a striking difference compared to Hamiltonian of the fully
diffeomorphism-invariant theory \eqref{HDUG}, which is a sum of first
class constraints. However, there exists a clear relation between these
Hamiltonians, since the nonvanishing term can be eliminated (or
introduced) via a simple time-dependent canonical transformation.
% \footnote{The following argument could be made just as well after the
% variable $\lambda$ has been split into a zero mode $\lambda_0$ and an
% average free mode $\bar\lambda$. That is we could establish the
% relation of the Hamiltonians \eqref{HDUG3} and \eqref{HUG3} in the
% exact same way.}

Consider the following two canonical transformations of the variable
$p_\lambda\rightarrow p'_\lambda$,
\begin{equation}\label{plambda.trans}
 p_\lambda=p'_\lambda\pm \epsilon_0t,
\end{equation}
with all other variables remaining unchanged. These two transformations
are generated by the functionals
\begin{equation}
 F_\pm=\int_{\Sigma_t}d^3x\left( \lambda p'_\lambda
 \pm \epsilon_0\lambda t \right),
\end{equation}
respectively. The Hamiltonian transforms to
\begin{equation}
 H'=H+\frac{\partial F_\pm}{\partial t}
 =H\pm\int_{\Sigma_t}d^3x\epsilon_0\lambda.
\end{equation}
We can see that the transformation generated by $F_+$ eliminates the
nonvanishing term from the Hamiltonian \eqref{HUG}, while the
transformation generated by $F_-$ introduces the nonvanishing term into
the Hamiltonian \eqref{HDUG}. Notice that the variable $p_\lambda$
appears only in the primary constraints $\cC_\lambda\approx0$ and
$p_\lambda\approx0$ of the two theories, and these constraints drop out
of the Hamiltonian due to the consistency conditions for their Lagrange
multipliers \eqref{vlambda} and \eqref{vlambda2}, respectively.

The theory with fixed metric determinant can be shown to be a
(partially) gauge fixed version of the fully diffeomorphism-invariant
theory. When we introduce the following gauge fixing conditions into the
Hamiltonian \eqref{HDUG3},\footnote{The first two gauge conditions are
associated with the first class constraints $\pi_N\approx0$ and
$\cH_0\approx0$ (the zero mode of $\cH_T\approx0$), respectively.}
\begin{equation}\label{unimodGauge}
 \cU=N\sqrt{h}-\epsilon_0\approx0,\quad p_\lambda^0\approx0,\quad
 \bar{p}_\lambda\approx0,\quad V^i\approx0,
\end{equation}
and together with the second class constraints $\cC_\lambda\approx0$ and
$p_{\bn}\approx0$, we obtain a Hamiltonian that has the same form as
\eqref{HUG3}, except for the extra nonvanishing term in \eqref{HUG3},
$-\int_{\Sigma_t}d^3x\epsilon_0\lambda_0$ with $\lambda_0$ given in
\eqref{lambda0.fixed}. That extra term can be introduced with the
canonical transformation \eqref{plambda.trans} of the variable
$p_\lambda$. Thus the theory \eqref{SUG} is a (partially) gauge fixed
version of the theory \eqref{SDUG}. In other words, the fully
diffeomorphism-invariant theory defined in \eqref{SDUG} (and analyzed
in Sect.~\ref{sec3}) is a generalization of the unimodular theory of
gravity with an enlarged gauge symmetry.

\section{Conclusions}\label{sec6}
We have studied path integral quantization of two versions of unimodular
gravity. In the fully diffeomorphism-invariant theory defined by the
action \eqref{SDUG}, the path integral has the same form as the one of
GR with a cosmological constant $\Lambda$ \eqref{ZDUGfinal}, except that
the value of $\Lambda$ is not set by the action. The cosmological
constant $\Lambda$ is an unspecified value of the variable $\lambda$.
There exist two approaches regarding the interpretation of $\Lambda$ in
this theory:
\begin{enumerate}[(i)]
\item The value of $\Lambda$ can be set in the boundary conditions of
the path integral, since it is a boundary value of the variable
$\lambda$. In this case, the value of $\Lambda$ is completely
unspecified by the theory, and hence it needs to be set to the desired
value by hand. One can use anthropic arguments for limiting the range of
possible values of $\Lambda$ (see \cite{Weinberg:1988cp,
Padmanabhan:2002ji,Nobbenhuis:2004wn,Bousso:2007gp,Burgess:2013ara}
for reviews), but we do not consider such arguments here.
Physically, it makes no difference whether the observed value of
$\Lambda$ is fixed by the boundary conditions or by setting the value
of a coupling constant in the Lagrangian. Thus this approach is
physically equivalent to GR.

\item Since the value of $\Lambda$ is unspecified, the vacuum state of
the universe can be defined as a superposition of vacuum states
corresponding to different values of $\Lambda$ \cite{Weinberg:1988cp}.
Such an approach was used in \cite{Ng:1990rw,Ng:1990xz} where the path
integral of the form \eqref{ZNvD} was conjectured. A similar path
integral was later obtained in \cite{Smolin:2009ti}. Starting from
the action \eqref{SDUG}, we have derived the path integral
\eqref{ZNvD} without any addition or manipulation of variables. The
integration over $\Lambda$ arises due to the definition of the vacuum
state \eqref{vacOmega}. The given theory shows that it is unnecessary
to impose the unimodular condition on the metric determinant in order
to obtain the path integral \eqref{ZNvD}.

Using the semiclassical approximation and the stationary phase
approximation one can argue \cite{Smolin:2009ti} that the path integral
\eqref{ZNvD} is dominated by the values of $\Lambda$ around the average
energy density of matter over spacetime \eqref{Lambda.dom}. It is
presumed that the given values of $\Lambda$ were included in the vacuum
state \eqref{vacOmega}. This result is interesting but problematic. In
order to estimate the average energy density of matter over spacetime,
we need information on both the matter and gravitational (background)
fields, which depend on the assumed value of $\Lambda$.
It could be interesting to search for alternative mechanisms that would
single out the most likely values of $\Lambda$ within the fully
diffeomorphism-invariant theory.
\end{enumerate}
\vspace{.5em}

In the more conventional case defined by the action \eqref{SUG}, the
path integral \eqref{ZUG} differs from the path integral of GR in two
ways: (i) since the zero mode of the super-Hamiltonian \eqref{cH0} is a
second class constraint, the first class Hamiltonian constraint
\eqref{bcH_T} and an associated gauge condition have zero average over
space, and (ii) the metric in the path integral must satisfy the
integrated unimodular condition \eqref{unimodcond.av}.
The condition \eqref{unimodcond.av} imposes the unimodular condition
\eqref{unimodcond} to hold in average over space at each moment in time.
The path integral has a generally noncovariant form due to the given
differences. The perturbative gravitational field in the (semiclassical)
quantum effective action must satisfy the integrated condition
\eqref{trf.av}.

At quantum level the unimodular condition can manifest itself in three
ways. In the HT theory \cite{Henneaux:1989zc}, the unimodular condition
is imposed locally in the path integral and in the quantum effective
action \cite{Smolin:2009ti}. In the path integral and the quantum
effective action of the UG theory \eqref{SUG}, the unimodular condition
is averaged over space \eqref{unimodcond.av}. Lastly, the DUG theory
\eqref{SDUG} does not involve a unimodular condition.

In Sect.~\ref{sec5}, we established the canonical relation of the two
considered versions of unimodular gravity. While the actions \eqref{SUG}
and \eqref{SDUG} are shown to be equivalent classically, the
time-dependent canonical transformation \eqref{plambda.trans} involved
in the relation of their Hamiltonian structures has an interesting
effect to the quantum theory. That is the appearance of the averaged
unimodular condition \eqref{unimodcond.av} in the path integral of UG
\eqref{ZUG.Ug}. Furthermore, the gauge symmetry is restricted, since
the integral of the super-Hamiltonian over space \eqref{cH0} becomes a
second class constraint. This implies that the path integral involves a
pair of ghost fields and a Lagrange multiplier field whose average
values over space must vanish.
In practice, both of these implications are inconvenient to work with.
Thus the fully diffeomorphism-invariant theory considered in
Sect.~\ref{sec3}, or the previously worked out HT theory, are the
preferable versions of unimodular gravity for quantization.

The differences in the path integrals of different versions of
unimodular gravity do not necessarily imply that the physical
predictions of the theories are different. The DUG and HT theories can
indeed be expected to be physically equivalent, since the theories are
related by a simple change of an auxiliary variable (see below
\eqref{SDUG}). However, in addition to gauge fixing, the canonical
relation between DUG and UG involves the time-dependent canonical
transformation \eqref{plambda.trans}, which leads to the aforementioned
complications. Therefore it is still unclear whether the path integrals
\eqref{ZDUGfinal} and \eqref{ZUG.Ug} produce equivalent predictions.
Confirming this would require the formulation of Feynman rules and the
calculation of the scattering matrices. This is a very demanding task in
itself, which we wish to investigate in further work.

Proper quantization of gravity requires more advanced methods. Two known
approaches are the spin foam models and the dynamical triangulations.
Some steps toward loop quantization of unimodular gravity have already
been taken in \cite{Bombelli:1991jj}, and more recently in
\cite{Smolin:2010iq}.

\subsection*{Acknowledgments}
We are grateful to M. Chaichian for discussions and suggestions.
R.B. thanks FAPESP for full support, Project No. 2013/26571-4.
M.O. thankfully acknowledges support from the Emil Aaltonen Foundation.
The support of the Academy of Finland under the Projects No. 136539 and
272919 is gratefully acknowledged.

\appendix
\section{Appendix: Quantization of gauge theories with linearly
dependent generators}\label{appendix}
The unimodular theories of gravity involve certain local gauge
generators whose integrals over the spatial hypersurface vanish by
definition. This type of a spatially nonlocal linear dependence of
generators is an inherent feature of the unimodular gravity theories,
where the cosmological constant appears as a (constant) value of a
scalar variable. Quantization of gauge theories with linearly dependent
generators was achieved in \cite{Batalin:1984jr}. The Batalin-Vilkovisky
formalism \cite{Batalin:1984jr} is suitable for the description of the
nonlocally linearly dependent generators of unimodular gravity. In
particular, the average free nature of the Faddeev-Popov ghosts and
auxiliary fields associated with the average free generators is
explained naturally within the given formalism.

First we review the Batalin-Vilkovisky formalism. The formalism was
applied to the minisuperspace formulation of Friedman-Robertson-Walker
cosmology models in \cite{Barvinsky:2010yx}, where a review of the
formalism for theories with only bosonic gauge fields is also presented.
Since the gravitational sectors of the unimodular gravity theories
involve only bosonic fields, our presentation follows
\cite{Barvinsky:2010yx} with a few conventional differences due to
the following application to the unimodular gravity theories
considered in sections~\ref{sec3} and \ref{sec4}.

When the generators $G_\alpha$ are linearly dependent, there
exist right zero eigenvectors $Z^\alpha_a$,
\begin{equation}\label{G.dependent}
 G_\alpha Z^\alpha_a=0.
\end{equation}
Here the condensed index $\alpha$ labels each local generator at every
point in the spatial hypersurfaces. Hence summing over such an index
involves an integration over space in addition to a sum over the
components.\footnote{Unlike in \cite{Barvinsky:2010yx} the sum over a
condensed index does not involve integration over time. Furthermore we
do not consider Euclidean quantum gravity, i.e., Wick rotation of
time is not performed.}
The Latin index labels the zero eigenvectors $a=1,\ldots,A$.
Here the vectors $Z^\alpha_a$ are linearly independent, i.e., we
consider a first-stage reducible theory.
The gauge conditions $\chi^\alpha$ have to be similarly redundant as
the generators, so that there exists left zero eigenvectors
$\hat{Z}_\alpha^a$,
\begin{equation}\label{chi.dependent}
 \hat{Z}_\alpha^a\chi^\alpha=0.
\end{equation}
The eigenvectors $Z^\alpha_a$ and $\hat{Z}_\alpha^a$ are the right and
left zero vectors of the degenerate Faddeev-Popov operator,
\begin{equation}
 Q^\alpha_{\ \beta}=\pb{\chi^\alpha,G_\beta},
\end{equation}
respectively. Thus, within this formalism, the Faddeev-Popov ghosts
$c^\alpha$, $c^{*}_\alpha$ become gauge fields that require additional
gauge fixing. For that purpose the set of Lagrange multipliers and
ghosts ($c^\alpha$, $c^{*}_\alpha$, $ \eta_\alpha$) is extended to
\cite{Batalin:1984jr}
\begin{equation}
 \Phi_\mathrm{g}=\left( c^\alpha, c^{*}_\alpha, \eta_\alpha, C^a,
 C^{*}_a, E^a, \theta_a, \vartheta^a \right),
\end{equation}
where $c^\alpha$, $c^{*}_\alpha$, $\theta_a$, $\vartheta^a$ are
Grassmann anti-commuting variables and the rest are commuting variables.
The path integral and the corresponding effective gauge fixed action are
written as
\begin{equation}\label{Zgld}
\begin{split}
 Z&=\int\cD\phi^i\cD\pi_i\cD\Phi_\mathrm{g}\exp\left(
 \frac{i}{\hbar} S_\mathrm{eff} \right),\\
 S_\mathrm{eff}&=S -\int dt \big[
 c^{*}_\alpha Q^\alpha_{\ \beta}c^\beta
 +C^{*}_a(\omega^a_\alpha Z^\alpha_b)C^b
 +\eta_\alpha(\chi^\alpha+\sigma^\alpha_aE^a) \\
 &\quad +\theta_a\omega^a_\alpha c^\alpha
 +c^{*}_\alpha\sigma^\alpha_a\vartheta^a \big],
\end{split}
\end{equation}
where $\phi^i$ and $\pi_ i$ are the gauge fields and their canonically
conjugated momenta, and $S$ is the action without gauge fixing. The
extra Lagrange multipliers $(\theta_a$, $\vartheta^a)$ impose the gauge
conditions $\omega^a_\alpha c^\alpha$ and $c^{*}_\alpha\sigma^\alpha_a$
on the Faddeev-Popov ghosts, where the gauge parameters
$(\omega^a_\alpha$, $\sigma^\alpha_a)$ are arbitrary. The variables
$C^{*}_a$ and $C^a$ are the ghosts for the Faddeev-Popov ghost fields.
The so-called extra ghosts $E^a$ regulate divergent factors $\delta(0)$
that appear in the original gauge fixing $\delta(\chi^\alpha)$ with a
redundant set of gauge conditions \eqref{chi.dependent}.

Integration over the ghost sector gives the path integral as
\begin{equation}\label{Zgld2}
 Z=\int\cD\phi^i\cD\pi_i
 \frac{\det\cF^\alpha_{\ \beta}}{\det q^a_b \det\hat{q}^a_b}
 \int\cD E^a\delta(\chi^\alpha+\sigma^\alpha_aE^a) (\det \hat{q}^a_b)
 \exp\left(\frac{i}{\hbar}S\right),
\end{equation}
where the gauge fixed Faddeev-Popov operator is defined as
\begin{equation}\label{FPop.gf}
 \cF^\alpha_{\ \beta}=Q^\alpha_{\ \beta}+\sigma^\alpha_a
 \omega^a_\beta,
\end{equation}
and the following matrices are introduced
\begin{equation}\label{qmatrices}
 q^a_b=\omega^a_\alpha Z^\alpha_b,\quad
 \hat{q}^a_b=\hat{Z}^a_\alpha\sigma^\alpha_b.
\end{equation}
The path integral \eqref{Zgld2} is independent of the chosen gauge
parameters $(\omega^a_\alpha$, $\sigma^\alpha_a)$, since both the
ratio of determinants $(\det\cF^\alpha_{\ \beta}/\det q^a_b
\det\hat{q}^a_b)$ and the regulated gauge fixing factor are
invariant under a change of the gauge parameters
(see \cite{Barvinsky:2010yx} for a proof).

Next we apply this formalism to the quantization of the two unimodular
gravity theories (DUG and UG).

\subsection{Fully diffeomorphism-invariant unimodular gravity}
Let us consider the quantization of DUG presented in Sect.~\ref{sec3}.
The second class constraints are given in \eqref{scc.DUG}.
The generators are
\begin{equation}
 G_\alpha=\left[ \pi_N,\pi_i,p_i,\cH'_T,\cH_i,\bar{\cC}' \right]
\end{equation}
with \eqref{fcc.DUG}.
Gauge fixing conditions are chosen as in \eqref{gauge.GRlike},
\begin{equation}
 \chi^\alpha=\left[ \sigma^0,\sigma^i,V^i,\chi^0,\chi^i,\bar{p}_\lambda
 \right].
\end{equation}
The generator $\bar{\cC}'$ and the corresponding gauge condition
$\bar{p}_\lambda$ exhibit a nonlocal linear dependence, since
their integrals over space vanish by definition. Hence there exist a
single right zero vector,
\begin{equation}\label{rzv.DUG}
\begin{split}
 Z^\alpha &=\left[0,0,0,0,0,\frac{\sqrt{h}}{\int_{\Sigma_t}d^3x\sqrt{h}}
 \right],\\
 G_\alpha Z^\alpha &=\frac{1}{\int_{\Sigma_t}d^3x\sqrt{h}}
 \int_{\Sigma_t}d^3x\sqrt{h}\bar{\cC}'=0,
\end{split}
\end{equation}
and a single left zero vector as
\begin{equation}\label{lzv.DUG}
\begin{split}
 \hat{Z}_\alpha &=\left[0,0,0,0,0,1\right],\\
 \hat{Z}_\alpha\chi^\alpha &=\int_{\Sigma_t}d^3x\bar{p}_\lambda=0.
\end{split}
\end{equation}
Since only one pair of zero vectors exist, we have dropped the label
$a$ from the zero vectors and also from the other variables of the
path integral \eqref{Zgld}.

We choose the gauge fixing parameters for the ghosts as
\begin{equation}\label{gfparam.DUG}
\begin{split}
 \omega_\alpha&=\left[0,0,0,0,0,-1\right],\\
 \sigma^\alpha&=\left[0,0,0,0,0,\frac{\sqrt{h}}
 {\int_{\Sigma_t}d^3x\sqrt{h}}\right].
\end{split}
\end{equation}
Thus the ghost fields $(\bar{c},\bar{c}^{*})$ associated with the
generator $\bar{\cC}'$ are imposed to satisfy the gauge conditions
\begin{equation}\label{gf.ghosts.DUG}
\begin{split}
 \omega_\alpha c^\alpha &=-\int_{\Sigma_t}d^3x\bar{c}=0,\\
 c^{*}_\alpha\sigma^\alpha &=\frac{1}{\int_{\Sigma_t}d^3x\sqrt{h}}
 \int_{\Sigma_t}d^3x\sqrt{h}\bar{c}^{*}=0.
\end{split}
\end{equation}
We obtain the (now one-dimensional) matrices \eqref{qmatrices} as
\begin{equation}
 q=\omega_\alpha Z^\alpha=-1,\quad
 \hat{q}=\hat{Z}_\alpha\sigma^\alpha=1.
\end{equation}
In the amended Faddeev-Popov operator \eqref{FPop.gf}, the gauge fixing
term with \eqref{gfparam.DUG} contributes to the part of
$c^{*}_\alpha\cF^\alpha_{\ \beta}c^\beta$ that involves the ghosts
$\bar{c},\bar{c}^{*}$ as
\begin{equation}
 \int_{\Sigma_t}d^3xd^3y\, \bar{c}^{*}(x)\pb{\bar{p}_\lambda(x),
 \bar{\cC}'(y)} \bar{c}(y)
 +c^{*}_\alpha\sigma^\alpha\omega_\beta c^\beta
 =-\int_{\Sigma_t}d^3x\bar{c}^{*}\bar{c}.
\end{equation}
This implies a unit contribution to the canonical measure of the
path integral.

The gauge condition $\bar{p}_\lambda=0$ imposes $p_\lambda$ to become
proportional to a spatial constant $p_\lambda^0$ \eqref{lambdadec},
which is the integrated value of $p_\lambda$ over space
\eqref{lambdazero}.
The extra ghost $E$ introduces an independent term into this gauge
condition, so that the integral of the condition over space no longer
vanishes,
\begin{equation}
 \hat{Z}_\alpha\left( \chi^\alpha+\sigma^\alpha E \right)=E,
\end{equation}
which serves as a regulator for the corresponding $\delta$-function in
the path integral. The functional integral over $E$ forces the Lagrange
multiplier of the gauge condition $\bar{p}_\lambda$ to have vanishing
average value over space,
\begin{multline}
 \int\cD\eta\cD E\exp\left[ \frac{i}{\hbar}\int_{\cM}d^4x
 \eta\left( \bar{p}_\lambda
 +\frac{\sqrt{h}} {\int_{\Sigma_t}d^3x\sqrt{h}}E
 \right) \right]\\
 \propto\int\cD\eta \delta\left( \frac{\int_{\Sigma_t}d^3x\sqrt{h}\eta}
 {\int_{\Sigma_t}d^3x\sqrt{h}} \right)
 \exp\left( \frac{i}{\hbar}\int_{\cM}d^4x \eta\bar{p}_\lambda \right),
\end{multline}
which prevents the appearance of divergent factors $\delta(0)$.
In sections~\ref{sec3.3} and \ref{sec4.3}, every $\delta$-function for
an average free constraint is regulated in this way, and we denote such
$\delta$-functions simply as
\begin{equation}
 \int\cD\bar\eta\exp\left( \frac{i}{\hbar}\int_{\cM}d^4x
 \bar\eta\bar{p}_\lambda \right)=\delta(\bar{p}_\lambda),
\end{equation}
where the auxiliary field $\bar\eta$ is now assumed to have vanishing
average over space.

Once the additional gauge fixing \eqref{gf.ghosts.DUG} on the ghosts
associated with the linearly dependent generators and gauge conditions
is performed, and the $\delta$-functions of average free constraints are
regulated, it is easy see how the canonical path integral is obtained as
\eqref{ZDUG.first} after the nonphysical variables in the canonical
integration measure \eqref{PImeasure.DUG} have been integrated out
(except for those variables deleted by the gauge conditions $\chi^\mu$
that are unspecified). In summary, the path integral obtained in
Sect.~\ref{sec3.3} corresponds to the specific choice of the gauge
parameters \eqref{gfparam.DUG}, which are responsible for the
additional gauge fixing required by the linearly dependent generators.

\subsection{Unimodular gravity with a fixed metric determinant}
Here we consider the quantization of UG presented in Sect.~\ref{sec4}.
The generators are
\begin{equation}
 G_\alpha=\left[ \pi_i,\bcH'_T,\cH'_i \right]
\end{equation}
with \eqref{fcc.UG}. Gauge conditions are chosen as in \eqref{gauge.UG},
\begin{equation}
 \chi^\alpha=\left[ \sigma^i,\bar\chi^0,\chi^i \right].
\end{equation}
We again have a nonlocal linear dependence, since the integrals of
$\bcH'_T$ and $\bar\chi^0$ over the spatial hypersurface vanish.
A single pair of zero vectors is obtained as
\begin{equation}
\begin{split}
 Z^\alpha &=\left[ 0,1,0 \right],\quad
 Z^\alpha G_\alpha=\int_{\Sigma_t}d^3x\bcH'_T=0, \\
 \hat{Z}_\alpha &=\left[ 0,1,0 \right],\quad
 \hat{Z}_\alpha\chi^\alpha=\int_{\Sigma_t}d^3x\bar\chi^0=0.
\end{split}
\end{equation}

The gauge fixing parameters $\omega_\alpha$ and $\sigma^\alpha$ can be
chosen so that the ghosts $(\bar{c},\bar{c}^{*})$ associated with
the generator $\bcH'_T$ satisfy the condition of vanishing average
value over space. The parameters are chosen as
\begin{equation}\label{gfparam.UG}
\begin{split}
 \omega_\alpha &=\left[0,\frac{\sqrt{h}}{\int_{\Sigma_t}d^3x\sqrt{h}},
 0\right],\\
 \sigma^\alpha &=\left[0,\frac{\sqrt{h}}{\int_{\Sigma_t}d^3x\sqrt{h}},
 0\right],\\
\end{split}
\end{equation}
and the gauge conditions on the ghosts read as
\begin{equation}\label{gf.ghosts.UG}
\begin{split}
 \omega_\alpha c^\alpha &=\frac{1}{\int_{\Sigma_t}d^3x\sqrt{h}}
 \int_{\Sigma_t}d^3x\sqrt{h}\bar{c}=0,\\
 c^{*}_\alpha\sigma^\alpha &=\frac{1}{\int_{\Sigma_t}d^3x\sqrt{h}}
 \int_{\Sigma_t}d^3x\sqrt{h}\bar{c}^{*}=0.
\end{split}
\end{equation}
The determinants of the matrices \eqref{qmatrices} have unit values
\begin{equation}
 q=\omega_\alpha Z^\alpha=1,\quad
 \hat{q}=\hat{Z}_\alpha\sigma^\alpha=1.
\end{equation}

The second class constraints \eqref{scc.UG} contain a pair of average
free constraints $(\bar\lambda\approx0$, $\bar{p}_\lambda\approx0)$,
which have to be treated in a similar way as a nonlocally linear
dependent generator and a gauge condition. In the path integral,
the contribution of these constraints is just a unit factor to the
Faddeev-Popov determinant, which is quite similar to the case
of the constraints $\bar{\cC}'$ and $\bar{p}_\lambda$ in DUG.
Hence we shall omit the analysis of these constraints here.

Lastly, we explain how the path integral in Sect.~\ref{sec4.3} is
obtained from the present formalism. As was discussed above, all the
$\delta$-functions for average free constraints has to be regulated in
order to avoid divergent $\delta(0)$ factors. Integration over the
extra ghost $E$ and the additional Lagrange multipliers
$(\theta,\vartheta)$ produces the $\delta$-functions that impose the
Lagrange multiplier of the gauge condition $\bar\chi^0$ and the ghosts
$(\bar{c},\bar{c}^{*})$ associated with the linearly dependent
generators to become average free over the spatial hypersurface,
\begin{equation}\label{avfree.deltaf}
 \delta(\eta_\alpha\sigma^\alpha)\delta(\omega_\alpha c^\alpha)
 \delta(c^{*}_\alpha\sigma^\alpha),
\end{equation}
where the conditions for the ghosts are \eqref{gf.ghosts.UG} and the
condition for the Lagrange multiplier $\bar\eta$ is
\begin{equation}\label{gf.Lm.UG}
 \eta_\alpha\sigma^\alpha=\frac{1}{\int_{\Sigma_t}d^3x\sqrt{h}}
 \int_{\Sigma_t}d^3x\sqrt{h}\bar\eta=0.
\end{equation}
In Sect.~\ref{sec4.3}, the $\delta$-functions \eqref{avfree.deltaf}
are omitted in the path integral \eqref{ZUG.Ug}, since the fields
$\bar\chi^0$, $\bar{c}$, $\bar{c}^{*}$ are assumed to satisfy the
conditions \eqref{gf.ghosts.UG} and \eqref{gf.Lm.UG} from the beginning.
The Faddeev-Popov determinant in the path integral is defined in terms
of ghost fields that satisfy the conditions \eqref{gf.ghosts.UG}, so
that no zero modes are present in the ghost sector that generates the
determinant. %, $c^{*}_\alpha Q^\alpha_{\ \beta}c^\beta$.
In particular, the nontrivial part of the determinant in the path
integral \eqref{ZUG.first} (and thereafter) is written as\footnote{Note
that here the sums over the repeated indices $\mu,\nu$ include
integration over the spatial hypersurface.}
\begin{equation}
 \left|\det\pb{\tilde\chi^\mu,\tcH_\nu}\right|=\int\cD\tilde{c}^{*}_\mu
 \cD\tilde{c}^\mu \exp\left( -\frac{i}{\hbar}\int dt \tilde{c}^{*}_\mu
 \pb{\tilde\chi^\mu,\tcH_\mu}\tilde{c}^\nu \right),
\end{equation}
where the first components of the ghost fields $\tilde{c}^\mu=(\bar{c},
c^i)$ and $\tilde{c}^{*}_\mu=(\bar{c}^{*}, c^{*}_i)$ are assumed to be
average free over space. Finally, we can see that the path integral
obtained in Sect.~\ref{sec4.3} corresponds to the specific choice of
the gauge parameters \eqref{gfparam.UG}, which are responsible for
imposing the necessary conditions on the ghosts \eqref{gf.ghosts.UG} and
on the Lagrange multipliers \eqref{gf.Lm.UG}.

\end{document}